\RequirePackage[2020-02-02]{latexrelease}
\makeatletter\ifdefined\@@eqno\else\let\@@eqno\eqno\fi\makeatother
\RequirePackage{amsmath}[=2018-12-01]
\documentclass{rQUF2e}

\usepackage{array,booktabs}
\usepackage{url}

\graphicspath{{figures/}{../figures/}}

\theoremstyle{plain}
\newtheorem{theorem}{Theorem}[section]
\newtheorem{proposition}[theorem]{Proposition}
\newtheorem{lemma}[theorem]{Lemma}
\newtheorem{corollary}[theorem]{Corollary}
\newtheorem{conjecture}[theorem]{Conjecture}
\theoremstyle{definition}

\theoremstyle{remark}
\newtheorem{remark}[theorem]{Remark}

\newcommand{\D}{\mathrm{d}}
\newcommand{\E}{\mathbb{E}}
\newcommand{\Q}{\mathbb{Q}}
\renewcommand{\P}{\mathbb{P}}
\newcommand{\F}{\mathcal{F}}
\newcommand{\one}{\mathbf{1}}
\newcommand{\VIX}{\mathrm{VIX}}
\let\draftnote\empty

\begin{document}

\title{Rough log-normal stochastic volatility with quadratic drift: second-moment stability and the martingale property under positive correlation}
\author{Artur Sepp$^{\ast\dag}$\thanks{$^\ast$Corresponding author. Email: artursepp@gmail.com} and Parviz Rakhmonov$^{\ddag}$\\
\affil{$\dag$LGT Bank (Schweiz) AG, Z\"urich, Switzerland\\
$\ddag$Marex, London, United Kingdom}
\received{Manuscript word count 10,149}}
\maketitle
\thispagestyle{empty}

\begin{abstract}
We extend the log-normal stochastic volatility model with quadratic drift to a fractional Volterra equation. The volatility has a unique global non-negative solution with finite moments on every finite horizon. With linear drift we derive the conditional first and second moments, expected integrated variance and conditional squared VIX. We give the exact threshold for a uniformly bounded second moment and the exponential growth rate above it. With positive quadratic damping the second moment is uniformly bounded for every admissible Hurst exponent and volatility of volatility. A separate bound on long-run expected integrated variance per unit time depends only on the drift parameters. Under the share measure the volatility beta offsets the quadratic damping. The price is a true martingale when damping covers beta. For a positive finite exponential mixture this condition is also necessary. The converse for a singular fractional kernel remains open. We illustrate calibration under positive correlation on observed Bitcoin and Ether option smiles and benchmark the simulation against exact moments. A marked order-flow construction relates multiplicative volatility noise to signed trade impacts arriving at a rate proportional to variance. A small state-dependent bias generates the quadratic drift. We establish a stopped diffusion limit for the constant kernel.
\end{abstract}

\begin{classcode}C63; G13\end{classcode}

\section{Introduction}\label{sec:intro}

Log-volatility of equity indices behaves like a fractional Brownian motion with Hurst exponent of order $0.1$ \citep{GatheralJaissonRosenbaum2018}. \citet{BayerFrizGatheral2016} use this observation in the rough Bergomi model, in which the instantaneous variance is the exponential of a Gaussian Volterra process. The model generates the power-law decay of the at-the-money skew with few parameters. \citet{ElEuchRosenbaum2019} introduce the rough Heston model and derive its characteristic function from a fractional Riccati equation, and \citet{AbiJaberLarssonPulido2019} place it in the class of affine Volterra processes. The quadratic rough Heston model \citep{GatheralJusselinRosenbaum2020} and the quintic Ornstein--Uhlenbeck model \citep{AbiJaberIllandLi2022} extend the class towards joint calibration to SPX and VIX smiles.

The log-normal stochastic volatility model with quadratic drift of \citet{SeppRakhmonov2023} addresses a different gap. Its volatility follows a diffusion with linear volatility-of-volatility and with drift $(\kappa_1+\kappa_2\sigma)(\theta-\sigma)$. The quadratic term keeps the price process a true martingale when returns and volatility are positively correlated, provided $\kappa_2\ge\beta$, where $\beta$ is the volatility beta. Positive spot-volatility correlation is frequent in Bitcoin options \citep{SeppRakhmonov2023}. Standard rough Bergomi cannot combine positive spot-volatility correlation with a true-martingale price under the usual pricing measure \citep{Gassiat2019}. In this paper we develop the rough extension of the log-normal model with quadratic drift.

The model admits an interpretation in terms of trading activity and
the persistence of its effects. Suppose that trades generate signed
volatility innovations of comparable size and arrive at a rate
proportional to the prevailing variance. Aggregating these innovations
then gives a noise amplitude proportional to the volatility level:
more active markets generate larger absolute volatility fluctuations.
The fractional kernel represents a broad range of reaction horizons,
assigning power-law weights to past innovations rather than imposing
a single exponential decay scale. We interpret the drift
$(\kappa_1+\kappa_2\sigma)(\theta-\sigma)$ as a stabilising response
whose strength increases with volatility. Above the reference level
$\theta$, the leading term $-\kappa_2\sigma^2$ provides progressively
stronger damping. In the proposed order-flow construction, this term
arises from a small downward bias in volatility innovations combined
with an arrival rate proportional to $\sigma^2$. Economically, such
a bias could represent inventory adjustment and liquidity
replenishment following a volatility spike. This stabilising channel
can coexist with positive price--volatility co-movement when the
correlation describes joint innovations, whereas the drift describes
their subsequent correction. These assumptions provide an economic
interpretation of the coefficients.

\subsection{Related work}\label{sec:literature}
Table~\ref{tab:rbergomi} compares the volatility state, drift and correlation of the three models, together with their martingale and second-moment properties.

Rough Bergomi specifies Gaussian log-variance \citep{BayerFrizGatheral2016}, rough Heston uses affine variance \citep{ElEuchRosenbaum2019}, and quadratic rough Heston takes variance to be a quadratic function of a Volterra state \citep{GatheralJusselinRosenbaum2020}. Here the state is volatility itself. At $H=1/2$ its diffusion is that of \citet{SeppRakhmonov2023}. \citet{SeppRakhmonov2026} use closure under changes of measure and the true-martingale property of the discounted asset as criteria for a robust volatility model. Section~\ref{sec:measure} examines both criteria for the Volterra extension. Under linear drift our moment equation is a special case of polynomial Volterra moment theory \citep{AbiJaberCuchieroPelizzariPulidoSvalutoFerro2024}. We evaluate the stability threshold and the growth rate explicitly. For the price, \citet{Gassiat2019} studies the rough Bergomi martingale problem. \citet{GuyonJeanninJourdain2026} give the stopped share-measure reduction and cover our linear-drift finite-kernel case. Quadratic drift requires control of the tilted equation.

The order-flow interpretation distinguishes the variable moved by a trade from the rate of trading. \citet{HagerHorstWagenhoferXu2026} derive rough Bergomi using i.i.d.\ centred impacts on log-volatility and a constant order rate. \citet{JaissonRosenbaum2016} and \citet{ElEuchFukasawaRosenbaum2018} derive rough Heston from a nearly unstable Hawkes market with one-sided variance impacts. Our proposed construction uses signed volatility impacts and a rate proportional to squared volatility, with a small state-dependent mark bias. The construction is related to the quadratic Hawkes mechanism of \citet{BlancDonierBouchaud2017} and \citet{DandapaniJusselinRosenbaum2021}, but the positive-state and zero-floor identifications require care.

\citet{DandapaniJusselinRosenbaum2021} derive a quadratic-Hawkes limit with a positive variance floor and a signed trend, and \citet{ChudasamaIyer2025} extend their framework to asymmetric buy and sell impacts. Appendix~\ref{sec:micro_relation} states the exact normalised equation and the limits of the floor-free analogy.
Appendix~\ref{sec:micro} separates the exact generator calculation from the open global convergence problem.

The numerical method uses a Markovian lift \citep{AbiJaberElEuch2019,AlfonsiKebaier2024} and a drift splitting in the sense of \citet{Shinozaki2026}. We keep linear drift with factor decay and advance the price using beginning-of-step volatility. We combine conditional Monte Carlo \citep{Willard1997,RomanoTouzi1997,McCrickerdPakkanen2018} with multilevel sampling \citep{Giles2008}. Appendix~\ref{app:simulation} gives the three flows and the Gaussian half-tilt used to price options.
\subsection{Contributions}\label{sec:contributions}
We establish closure of the quadratic drift under the two changes of measure used in pricing, extending the diffusion analysis of \citet{SeppRakhmonov2023} to Volterra volatility. Both changes preserve the memory kernel, the diffusion coefficients and the constant and linear terms of the volatility drift. A market price of risk $\bar\lambda_j\sigma_t$ for each Brownian driver $j=0,1$ shifts the quadratic damping from $\kappa_2$ under the physical measure to $\kappa_2^{\Q}=\kappa_2+\beta\bar\lambda_0+\varepsilon\bar\lambda_1$ under the pricing measure. Proposition~\ref{prop:closure} gives the conditions under which the shifted polynomial admits the original mean-reverting parametrisation. When $\kappa_2^{\Q}\ge0$, the change-of-measure density is a true martingale on every finite horizon (Corollary~\ref{cor:density}).

Taking the asset as numeraire in the stopped equation adds $\beta\sigma^2$ to the volatility drift. The remaining damping is $\kappa_2^{\Q}-\beta$. When it is non-negative the tilted volatility cannot explode, so the price is a true martingale even with positive spot-volatility correlation (Theorem~\ref{thm:martingale}). This condition is also necessary for positive finite exponential mixtures (Proposition~\ref{prop:finite_converse}). Necessity remains open for a singular fractional kernel. Together these results extend the closure and martingality criteria of \citet{SeppRakhmonov2026} to the rough model.

The volatility equation has a unique global non-negative strong solution for both the fractional kernel and its finite exponential approximations. Under linear drift we obtain exact conditional first and second moments, the threshold for a uniformly bounded second moment and the growth rate above it. With positive quadratic damping the fractional model has a uniformly bounded second moment for every $0<H\le1/2$. Its long-run variance-swap strike has a ceiling that depends only on the drift parameters.

We impose the martingale condition when fitting the model to Bitcoin and Ether option smiles at fixed roughness. Appendix~\ref{sec:micro} relates the multiplicative volatility noise and quadratic drift to signed trade impacts and proves a stopped diffusion limit for the constant kernel.
\section{Model}\label{sec:model}

\subsection{Dynamics}

We work under the pricing measure $\Q$ with zero interest rates, so that $S_t$ denotes the forward price. The log-normal Volterra stochastic volatility model is
\begin{equation}\label{eq:model}
\begin{split}
\frac{\D S_t}{S_t} &= \sigma_t\,\D W^{(0)}_t,\\
\sigma_t &= \sigma_0+\int_0^t K(t-s)\Big[(\kappa_1+\kappa_2\sigma_s)(\theta-\sigma_s)\,\D s+\beta\sigma_s\,\D W^{(0)}_s+\varepsilon\sigma_s\,\D W^{(1)}_s\Big],
\end{split}
\end{equation}
where $W^{(0)}$ and $W^{(1)}$ are independent Brownian motions. The parameters satisfy $\sigma_0>0$, $\theta>0$, $\kappa_1\ge0$, $\kappa_2\ge0$, $\beta\in\mathbb{R}$ and $\varepsilon>0$. We denote the total volatility-of-volatility and the spot-volatility correlation by
\begin{equation}\label{eq:vartheta}
\vartheta^2=\beta^2+\varepsilon^2,\qquad \rho=\beta/\vartheta,
\end{equation}
and we write $\vartheta\,\D W^{\sigma}_t=\beta\,\D W^{(0)}_t+\varepsilon\,\D W^{(1)}_t$ for the Brownian motion driving the volatility. The kernel is the Riemann--Liouville fractional kernel
\begin{equation}\label{eq:kernel}
K(t)=\frac{t^{H-1/2}}{\Gamma(H+1/2)},\qquad H\in(0,1/2],\qquad \alpha:=H+\tfrac12 .
\end{equation}
For $H=1/2$ the kernel is constant, $K\equiv1$, and \eqref{eq:model} is the model of \citet{SeppRakhmonov2023}. The drift has the polynomial form $(\kappa_1+\kappa_2\sigma)(\theta-\sigma)=\kappa_1\theta-(\kappa_1-\kappa_2\theta)\sigma-\kappa_2\sigma^2$, which we use in the measure changes of Section~\ref{sec:measure}.

The log-price $X_t=\ln S_t$ and the integrated variance $I_t$ are
\begin{equation}\label{eq:logprice}
X_t=X_0-\frac12 I_t+\int_0^t\sigma_s\,\D W^{(0)}_s,\qquad I_t=\int_0^t\sigma_s^2\,\D s .
\end{equation}

\subsection{Forward volatility curve}

The volatility \eqref{eq:model} is neither a Markov process nor a semimartingale when $H<1/2$. Following \citet{ViensZhang2019}, we use as state variable the forward volatility curve
\begin{equation}\label{eq:theta_curve}
\Theta^t_s=\sigma_0+\int_0^t K(s-r)\Big[(\kappa_1+\kappa_2\sigma_r)(\theta-\sigma_r)\,\D r+\vartheta\sigma_r\,\D W^\sigma_r\Big],\qquad s\ge t .
\end{equation}
The curve $s\mapsto\Theta^t_s$ is $\F_t$-measurable, satisfies $\Theta^t_t=\sigma_t$, and gives the flow property
\begin{equation}\label{eq:flow}
\sigma_s=\Theta^t_s+\int_t^s K(s-r)\Big[(\kappa_1+\kappa_2\sigma_r)(\theta-\sigma_r)\,\D r+\vartheta\sigma_r\,\D W^\sigma_r\Big],\qquad s\ge t .
\end{equation}
The forward curve is an adapted input to the future Volterra equation and is not in general the conditional mean $\E[\sigma_s\mid\F_t]$. When $t$ advances, the curve moves in the direction $s\mapsto K(s-t)$ on $[t,\infty)$. This direction enters every functional derivative of the path-dependent Feynman--Kac formula of \citet{ViensZhang2019}. For $K\equiv1$ the curve is flat, $\Theta^t_s=\sigma_t$, and the path-dependent equations reduce to classical partial differential equations.

\subsection{Markovian approximation}

The implementation replaces $K$ by a sum of exponentials
\begin{equation}\label{eq:kernel_N}
K^N(t)=\sum_{i=1}^N w_i\,e^{-x_i t},\qquad w_i>0,\ x_i\ge0,\qquad \bar w:=\sum_{i=1}^N w_i ,
\end{equation}
with nodes $x_i$ and weights $w_i$. Markovian approximations of this type go back to \citet{AbiJaberElEuch2019}, and \citet{AlfonsiKebaier2024} analyse their convergence for completely monotone kernels. The calibration at $H=0.30$ uses a kernel with $N=97$ nodes (Appendix~\ref{sec:calibration_design}), and for $H=1/2$ the kernel is $K\equiv1$, with $N=1$, $x_1=0$ and $w_1=1$. Table~\ref{tab:martingale} and Figure~\ref{fig:martingale} use $N=3$, with nodes and weights from the quadrature rule of \citet{BayerBreneis2023}, optimised for European options with maturity $T$. For $H=0.1$ and $T=1$ the rule gives $x=(0.0333,\,2.242,\,46.83)$ and $w=(0.555,\,1.111,\,6.086)$.

The quadrature rule underestimates the mass of $K^2$, on which the second moment depends (Section~\ref{sec:stability}).

\begin{lemma}[Factor representation]\label{lem:factors}
Let $\sigma^N$ solve \eqref{eq:model} with $K^N$ in place of $K$. Define $\sigma_{0,i}=\sigma_0/\bar w$ and let $\sigma^{(i)}$, $i=1,\dots,N$, solve
\begin{equation}\label{eq:factors}
\D\sigma^{(i)}_t=-x_i\big(\sigma^{(i)}_t-\sigma_{0,i}\big)\D t+(\kappa_1+\kappa_2\sigma^N_t)(\theta-\sigma^N_t)\,\D t+\vartheta\sigma^N_t\,\D W^\sigma_t,\qquad \sigma^{(i)}_0=\sigma_{0,i},
\end{equation}
with $\sigma^N_t=\sum_{i=1}^N w_i\sigma^{(i)}_t$. Then $\sigma^N$ solves the Volterra equation with kernel $K^N$, and its forward curve is
\begin{equation}\label{eq:theta_N}
\Theta^{t}_s=\sigma_0+\sum_{i=1}^N w_i\,e^{-x_i(s-t)}\big(\sigma^{(i)}_t-\sigma_{0,i}\big),\qquad s\ge t .
\end{equation}
\end{lemma}

Lemma~\ref{lem:factors} reduces the approximated model to an $N$-dimensional diffusion, and \eqref{eq:theta_N} expresses the forward curve as a linear function of the factor states. The proof is in Appendix~\ref{app:proofs}. Any choice $\sigma_{0,i}$ with $\sum_i w_i\sigma_{0,i}=\sigma_0$ gives the same $\sigma^N$, and the code uses the equal split $\sigma_{0,i}=\sigma_0/\bar w$.

\subsection{Well-posedness}

For $\kappa_2=0$, the drift and diffusion coefficients of \eqref{eq:model} are globally Lipschitz. Theorem~3.3 and Example~2.3(ii) of \citet{AbiJaberLarssonPulido2019} give a unique continuous strong solution for the fractional kernel, and their Lemma~3.1 gives finite-horizon moments. Their Theorem~3.6 and Example~3.7 give a non-negative solution, because the kernel is completely monotone, the boundary drift is $\kappa_1\theta\ge0$, and the diffusion coefficient vanishes at zero. For $\kappa_2>0$ the drift grows quadratically and is not globally Lipschitz, so these results do not apply directly. The next theorem covers the quadratic drift, the tilted drifts of Section~\ref{sec:measure} and the Markovian kernels \eqref{eq:kernel_N}.

\begin{theorem}[Global well-posedness with quadratic drift]\label{thm:wellposed}
Let $K$ be the fractional kernel \eqref{eq:kernel} with $H\in(0,1/2]$ or a Markovian kernel \eqref{eq:kernel_N}. Let $\sigma_0\ge0$, $c_0\ge0$, $c_1\in\mathbb R$, $d\ge0$ and $\vartheta\ge0$, and let $B$ be a Brownian motion. The equation
\begin{equation}\label{eq:sve_general}
\sigma_t=\sigma_0+\int_0^tK(t-s)\big[c_0+c_1\sigma_s-d\sigma_s^2\big]\D s+\int_0^tK(t-s)\vartheta\sigma_s\,\D B_s
\end{equation}
has a pathwise unique continuous strong solution on $[0,\infty)$. The solution is non-negative, and $\E\big[\sup_{t\le T}\sigma_t^p\big]<\infty$ for every $p\ge1$ and $T>0$.
\end{theorem}

The volatility equation in \eqref{eq:model} is \eqref{eq:sve_general} with $c_0=\kappa_1\theta$, $c_1=\kappa_2\theta-\kappa_1$ and $d=\kappa_2$. The proof is in Appendix~\ref{app:proofs}. The difficulty is that a stochastic Volterra equation has no It\^o formula for functions of the state, so the Lyapunov argument that controls a quadratic drift for a diffusion is not available. We replace it by a one-sided bound. A truncated equation has a non-negative solution by the results cited above. On the non-negative half-line the term $-d\sigma^2$ only lowers the drift, and the non-negative kernel carries this sign through the convolution. The moments of the truncated solutions are therefore bounded by those of a linear majorant, uniformly in the truncation level, and we remove the truncation.

\citet{PannierJacquier2019} remove a truncation in the same way for a locally Lipschitz drift, under a linear growth condition that excludes the quadratic term.

The moment bound of Theorem~\ref{thm:wellposed} holds on finite horizons and inherits the growth of the linear majorant. Section~\ref{sec:moments} quantifies the long-horizon behaviour. At $H=1/2$ the equation is the classical positive, non-explosive diffusion of \citet{SeppRakhmonov2023}. For $H<1/2$ the theorem gives $\sigma_t\ge0$ and leaves strict positivity open.

\begin{conjecture}[Strict positivity]\label{conj:wellposed}
For every $H\in(0,1/2)$ and the parameters of \eqref{eq:model}, the volatility is almost surely strictly positive on every finite time interval.
\end{conjecture}

No result of this paper uses Conjecture~\ref{conj:wellposed}. Neither zero-hitting paths of a numerical scheme nor the microscopic sign bias of Appendix~\ref{sec:micro_model} resolves it.

\section{Moments and the stability parameter}\label{sec:moments}

With $\kappa_2=0$ the drift is linear, and the volatility solves a linear stochastic Volterra equation with multiplicative noise. The first two conditional moments are then available in closed form up to the resolvent of the kernel. The economic content is simple. Mean reversion acts on the volatility through the kernel, and its cumulative effect is the resolvent. The noise acts through the same kernel filtered by the resolvent, and its second moment compounds through the singularity of the kernel at zero.

\subsection{Resolvent}

Let $R$ denote the resolvent of $\kappa_1K$ \citep[Chapter~2]{GripenbergLondenStaffans1990}, defined by
\begin{equation}\label{eq:resolvent}
R=\kappa_1K-\kappa_1K*R,\qquad (K*R)(t)=\int_0^tK(t-u)R(u)\,\D u,
\end{equation}
and define $G=R/\kappa_1$ for $\kappa_1>0$ and $G=K$ for $\kappa_1=0$. For the fractional kernel \eqref{eq:kernel} the resolvent is explicit \citep{ElEuchRosenbaum2019},
\begin{equation}\label{eq:resolvent_frac}
R(t)=\kappa_1t^{\alpha-1}E_{\alpha,\alpha}(-\kappa_1t^{\alpha}),\qquad \int_0^tR(u)\,\D u=1-E_\alpha(-\kappa_1t^\alpha),\qquad G(t)=t^{\alpha-1}E_{\alpha,\alpha}(-\kappa_1t^{\alpha}),
\end{equation}
where $E_{a,b}(z)=\sum_{k\ge0}z^k/\Gamma(ak+b)$ is the Mittag-Leffler function and $E_a=E_{a,1}$. For the Markovian kernel \eqref{eq:kernel_N} the resolvent follows from a linear system of $N$ ordinary differential equations.

\subsection{Conditional mean and second moment}

\begin{proposition}[Conditional mean]\label{prop:mean}
Let $\kappa_2=0$ and $t\le s$. The conditional mean $m_t(s)=\E[\sigma_s\mid\F_t]$ is
\begin{equation}\label{eq:cond_mean}
m_t(s)=\theta+\psi_t(s)-\int_t^sR(s-r)\,\psi_t(r)\,\D r,\qquad \psi_t(s)=\Theta^t_s-\theta .
\end{equation}
In particular, $m(t):=\E[\sigma_t]=\theta+(\sigma_0-\theta)\big(1-\int_0^tR(u)\,\D u\big)$, and for the fractional kernel $m(t)=\theta+(\sigma_0-\theta)E_\alpha(-\kappa_1t^\alpha)$.
\end{proposition}

\begin{proposition}[Conditional second moment]\label{prop:second}
Let $\kappa_2=0$ and $t\le s$. The conditional second moment $v_t(s)=\E[\sigma_s^2\mid\F_t]$ solves the linear Volterra equation
\begin{equation}\label{eq:cond_second}
v_t(s)=m_t(s)^2+\vartheta^2\int_t^sG(s-r)^2\,v_t(r)\,\D r .
\end{equation}
Equivalently, $v_t=m_t^2+\widetilde R*m_t^2$ on $[t,\infty)$, where $\widetilde R$ is the resolvent of $\vartheta^2G^2$ with positive sign, $\widetilde R=\vartheta^2G^2+\vartheta^2G^2*\widetilde R$.
\end{proposition}

The proofs are in Appendix~\ref{app:proofs}. The key step writes the centred volatility $\sigma_s-m_t(s)$ as the stochastic convolution $\int_t^sG(s-r)\vartheta\sigma_r\,\D W^\sigma_r$. The deterministic mean reversion is absorbed in the resolvent, and It\^o's isometry then closes the second moment. The cross-time covariances of the volatility never enter \eqref{eq:cond_second}, which is why the equation is linear and one-dimensional.

With $\kappa_2=0$ the drift is affine and the squared diffusion coefficient is quadratic in the state. The volatility is then a polynomial Volterra process in the sense of \citet{AbiJaberCuchieroPelizzariPulidoSvalutoFerro2024}, whose moments solve deterministic integral equations. Propositions~\ref{prop:mean} and~\ref{prop:second} give the first two of these equations in resolvent form.

\begin{corollary}[Mittag-Leffler growth]\label{cor:ml}
Let $\kappa_1=\kappa_2=0$ and let $K$ be the fractional kernel \eqref{eq:kernel}. Then
\begin{equation}\label{eq:ml_second}
\E[\sigma_t^2]=\sigma_0^2\,E_{2H}\big(a\,t^{2H}\big),\qquad a=\frac{\vartheta^2\,\Gamma(2H)}{\Gamma(H+1/2)^2},
\end{equation}
and $\E[I_T]=\sigma_0^2\,T\,E_{2H,2}\big(a\,T^{2H}\big)$. As $t\to\infty$, $\E[\sigma_t^2]\sim(2H)^{-1}\sigma_0^2\exp\big(a^{1/(2H)}t\big)$.
\end{corollary}

At $H=1/2$ the second moment is $\sigma_0^2e^{\vartheta^2t}$. For $H<1/2$ its exponential growth rate is $a^{1/(2H)}$. Comparisons across $H$ require a specified horizon because $\vartheta$ has units year$^{-H}$.
\subsection{Long-horizon second-moment stability}\label{sec:stability}

Equation \eqref{eq:cond_second} is a linear Volterra equation with the non-negative kernel $h=\vartheta^2G^2$. Its positive resolvent is $\widetilde R=\sum_{j\ge1}h^{*j}$, and Tonelli's theorem gives $\int_0^\infty\widetilde R=\sum_{j\ge1}\gamma^j$, where $\gamma=\int_0^\infty h$. The series is finite exactly when $\gamma<1$. Since the squared mean is bounded above and away from zero under the next proposition's hypotheses, the same threshold characterises uniform second-moment boundedness.

\begin{proposition}[Long-horizon second-moment stability]\label{prop:stationarity}
Let $\kappa_2=0$, $\kappa_1>0$, $\theta>0$ and $\sigma_0>0$, and define
\begin{equation}\label{eq:gamma}
\begin{split}
\gamma&:=\vartheta^2\int_0^\infty G(u)^2\,\D u,\\
\int_0^\infty G(u)^2\,\D u&=\frac1\pi\int_0^\infty\frac{\D\omega}{\omega^{2\alpha}+2\kappa_1\cos(\pi\alpha/2)\,\omega^\alpha+\kappa_1^2}=C(\alpha)\,\kappa_1^{-2H/(H+1/2)},
\end{split}
\end{equation}
where
\begin{equation*}
C(\alpha)=-\frac{\cot(\pi\alpha/2)}{\alpha\sin(\pi/\alpha)},\qquad \tfrac12<\alpha<1,\qquad C(1)=\tfrac12.
\end{equation*}
Then $\sup_{t\ge0}\E[\sigma_t^2]<\infty$ if and only if $\gamma<1$. If $\gamma<1$, then $\E[\sigma_t^2]\to\theta^2/(1-\gamma)$ as $t\to\infty$, and if $\gamma\ge1$, then $\E[\sigma_t^2]\to\infty$.
\end{proposition}

The proof is in Appendix~\ref{app:proofs}. At $H=1/2$ the condition reads $\vartheta^2<2\kappa_1$, and the limit $2\kappa_1\theta^2/(2\kappa_1-\vartheta^2)$ is the second moment of the inverse gamma stationary law of \citet{SeppRakhmonov2023}. For $\kappa_1=0$ the integral in \eqref{eq:gamma} diverges, which is consistent with Corollary~\ref{cor:ml}. We call $\gamma$ the linear-drift stability parameter. The parameter is also the squared feedback-kernel norm in the signed quadratic-Hawkes comparison in Appendix~\ref{sec:micro}. The criterion concerns the coefficients under the specified measure, and it does not extend unchanged to quadratic drift.

The criterion carries over to the Markovian kernel \eqref{eq:kernel_N} with positive nodes. The transform $\widehat K^N(z)=\sum_{i=1}^Nw_i/(z+x_i)$ gives the rational function $\widehat G_N=\widehat K^N/(1+\kappa_1\widehat K^N)$, which has simple negative poles and positive residues. The resolvent $G_N$ is therefore a finite sum of decaying exponentials with positive coefficients, and the mean moves monotonically from $\sigma_0$ to $m_\infty=(\sigma_0+\kappa_1k_\infty\theta)/(1+\kappa_1k_\infty)$, where $k_\infty=\sum_{i=1}^Nw_i/x_i$. The proof of Proposition~\ref{prop:stationarity} applies with $\gamma_N:=\vartheta^2\int_0^\infty G_N(u)^2\,\D u$ in place of $\gamma$ and with $m_\infty$ in place of $\theta$. The second moment is uniformly bounded if and only if $\gamma_N<1$, and its limit is then $m_\infty^2/(1-\gamma_N)$. The threshold depends on the kernel, and Section~\ref{sec:illustrations} compares a three-factor kernel with the fractional kernel.

Above the threshold the second moment grows exponentially, and its rate solves a scalar equation. Define the Laplace transform of the kernel $h=\vartheta^2G^2$,
\begin{equation}\label{eq:h_hat}
\widehat h(\lambda)=\vartheta^2\int_0^\infty e^{-\lambda u}G(u)^2\,\D u
=\frac{\vartheta^2}{\pi}\int_0^\infty\frac{\D\omega}{\big|(\lambda/2+\mathrm{i}\omega)^\alpha+\kappa_1\big|^2},\qquad\lambda>0 .
\end{equation}
The function $\widehat h$ is continuous and strictly decreasing, with $\widehat h(0+)=\gamma$ and $\widehat h(\lambda)\to0$ as $\lambda\to\infty$.

\begin{proposition}[Growth rate above the threshold]\label{prop:growth}
Let $\kappa_2=0$, $\kappa_1\ge0$, $\theta>0$ and $\sigma_0>0$, and set $\gamma=\infty$ when $\kappa_1=0$. If $\gamma>1$, the equation $\widehat h(\lambda^*)=1$ has a unique root $\lambda^*>0$, and
\begin{equation}\label{eq:growth}
\lim_{t\to\infty}e^{-\lambda^*t}\,\E[\sigma_t^2]=\frac{1}{|\widehat h'(\lambda^*)|}\int_0^\infty e^{-\lambda^*s}m(s)^2\,\D s .
\end{equation}
For $\sigma_0=\theta$ the limit equals $\theta^2/(\lambda^*|\widehat h'(\lambda^*)|)$. If $\gamma=1$, then $\E[\sigma_t^2]/t\to\theta^2/\mu_h$ as $t\to\infty$, where $\mu_h=\vartheta^2\int_0^\infty u\,G(u)^2\,\D u<\infty$.
\end{proposition}

The proof in Appendix~\ref{app:proofs} reads \eqref{eq:cond_second} as a renewal equation. For $\kappa_1=0$ the transform is $\widehat h(\lambda)=a\lambda^{-2H}$, so that $\lambda^*=a^{1/(2H)}$ and the limit is $\sigma_0^2/(2H)$, in agreement with Corollary~\ref{cor:ml}. At $H=1/2$ the root is $\lambda^*=\vartheta^2-2\kappa_1$. Mean reversion lowers the rate only moderately when the kernel is rough. At $H=0.1$ and $\vartheta=1.85$ the rate is $17{,}851$ per year for $\kappa_1=0$ and $17{,}235$ per year for $\kappa_1=3$.

Table~\ref{tab:critical} lists the critical volatility-of-volatility $\vartheta_c=(\int_0^\infty G^2)^{-1/2}$. The second moment is unbounded at and above this threshold. Mean reversion raises it as $\vartheta_c\propto\kappa_1^{H/\alpha}$, with exponent $1/6$ at $H=0.1$ and $1/2$ at $H=1/2$. A tenfold increase of $\kappa_1$ therefore raises $\vartheta_c$ by $47\%$ and $216\%$, respectively. Parameters have Hurst-dependent units, so comparisons across rows require a horizon and a quotation convention. Freezing the volatility coefficient over a short horizon $\Delta$ gives the leading-order innovation proxy
\begin{equation*}
\nu_\Delta=\vartheta\Big(\int_0^\Delta K(u)^2\,\D u\Big)^{1/2}
=\frac{\vartheta\Delta^H}{\sqrt{2H}\,\Gamma(\alpha)}.
\end{equation*}
The proxy reduces to $\vartheta\sqrt\Delta$ at $H=1/2$. The right block of Table~\ref{tab:critical} expresses the boundary using $\nu_{1/252,c}$. The proxy freezes the volatility coefficient over a short horizon and differs from the exact conditional daily standard deviation. The difference matters especially when the value exceeds one. Whether the linear-drift constraint binds in a market depends on the calibrated triple $(H,\kappa_1,\vartheta)$ with its stated units.

The proxy also links $\vartheta$ to the volatility-of-volatility $\eta_{\mathrm{B}}$ of rough Bergomi, in which the variance is $\xi_0(t)\,\mathcal{E}\big(\eta_{\mathrm{B}}\sqrt{2H}\int_0^t(t-s)^{H-1/2}\,\D W_s\big)$ \citep{BayerFrizGatheral2016}. Matching the short-time kernels of the log-volatility in the two models gives $\vartheta=\eta_{\mathrm{B}}\sqrt{2H}\,\Gamma(\alpha)/2$ and therefore $\nu_\Delta=(\eta_{\mathrm{B}}/2)\Delta^H$.

\subsection{Second moment under the quadratic drift}\label{sec:classical_qd}

The quadratic drift removes the linear-drift second-moment constraint when the kernel is constant. In the generator applied to $\sigma^2$, its contribution is $-2\kappa_2\sigma^3$, which dominates the quadratic diffusion contribution $\vartheta^2\sigma^2$ at large volatility for every $\vartheta$.

\begin{proposition}[Quadratic drift at $H=1/2$]\label{prop:classical_qd}
Let $K\equiv1$, $\kappa_1>0$, $\theta>0$ and $\kappa_2>0$. The stationary density of $\sigma$ is proportional to
\begin{equation}\label{eq:stationary_density}
\sigma^{-2-2(\kappa_1-\kappa_2\theta)/\vartheta^2}\exp\Big(-\frac{2\kappa_1\theta}{\vartheta^2\sigma}-\frac{2\kappa_2\sigma}{\vartheta^2}\Big),
\end{equation}
and every positive and negative moment under the stationary law is finite for every $\vartheta>0$. From any deterministic $\sigma_0>0$, the second moment is uniformly bounded on $[0,\infty)$.
\end{proposition}

The proof is in Appendix~\ref{app:proofs}.

The density \eqref{eq:stationary_density} is of generalised inverse Gaussian type, the steady-state law that \citet{CarrWillems2019} obtain for a log-normal volatility with quadratic drift.
With $\kappa_2=0$, \eqref{eq:stationary_density} reduces to the inverse gamma law of \citet{SeppRakhmonov2023}, whose second moment is finite exactly when $\vartheta^2<2\kappa_1$, or $\gamma<1$ at $H=1/2$. With $\kappa_2>0$, the exponential tail gives all stationary moments, and the Lyapunov estimate in the proof gives a uniform second-moment bound from a deterministic initial state.

For the fractional kernel the generator argument is not available, but the quadratic drift still bounds the second moment in time average. The volatility is non-negative, which forces a lower bound on the semimartingale that drives the Volterra equation. The expected quadratic damping accumulated along the path cannot exceed that bound. The argument uses only the non-negativity of the volatility and of the resolvent of the first kind of the kernel, and therefore it also covers $K^N$.

\begin{proposition}[Time-averaged second moment under quadratic drift]\label{prop:cesaro}
Let $K$ be the fractional kernel \eqref{eq:kernel} with $H\in(0,1/2]$ or the kernel $K^N$. Let $L$ be its resolvent of the first kind, the non-negative measure on $[0,\infty)$ with $(K*L)(t)=\int_{[0,t]}K(t-s)\,L(\D s)=1$ for all $t>0$, and let $\kappa_2>0$. For every $T>0$ the average $A_T=\E[I_T]/T$ satisfies
\begin{equation}\label{eq:cesaro}
\kappa_2A_T\le\kappa_1\theta+(\kappa_2\theta-\kappa_1)^+\sqrt{A_T}+\frac{\sigma_0\,L([0,T])}{T} .
\end{equation}
For the fractional kernel $L([0,T])/T=T^{-\alpha}/\Gamma(2-\alpha)$ and, for every $\vartheta>0$,
\begin{equation}\label{eq:cesaro_limit}
\limsup_{T\to\infty}\frac{\E[I_T]}{T}\le\theta\max\Big(\theta,\frac{\kappa_1}{\kappa_2}\Big) .
\end{equation}
For $K^N$ the ratio $L([0,T])/T$ converges to $\ell_N=\big(\int_0^\infty K^N(s)\,\D s\big)^{-1}=\big(\sum_{i=1}^Nw_i/x_i\big)^{-1}$, with $\ell_N=0$ if a node vanishes. With $c_1=\kappa_2\theta-\kappa_1$ and for every $\vartheta>0$,
\begin{equation}\label{eq:cesaro_limit_N}
\limsup_{T\to\infty}\frac{\E[I_T]}{T}\le\bigg(\frac{c_1^++\sqrt{(c_1^+)^2+4\kappa_2(\kappa_1\theta+\sigma_0\ell_N)}}{2\kappa_2}\bigg)^{2} .
\end{equation}
\end{proposition}

The proof is in Appendix~\ref{app:proofs}. The bound \eqref{eq:cesaro_limit} depends neither on $\vartheta$ nor on $H$. The bound \eqref{eq:cesaro_limit_N} does not depend on $\vartheta$ either. Its additional term $\sigma_0\ell_N$ arises because an integrable kernel pulls the volatility back to the initial level $\sigma_0$. With linear drift and $\gamma\ge1$ the same average diverges by Proposition~\ref{prop:stationarity}. At $H=1/2$ the stationary law of Proposition~\ref{prop:classical_qd} satisfies $\E[\sigma^2]=\theta\,\E[\sigma]+(\kappa_1/\kappa_2)(\theta-\E[\sigma])$ with $0\le\E[\sigma]\le\theta$, which is consistent with \eqref{eq:cesaro_limit}.

The long-run variance-swap strike is bounded by drift parameters alone. Figure~\ref{fig:martingale} compares the finite-horizon bounds for a three-factor kernel with simulated time averages. For the fractional kernel, comparison with an affine process also bounds the second moment at each time.

\begin{corollary}[Uniform second moment under quadratic drift]\label{cor:pointwise}
For the fractional kernel \eqref{eq:kernel}, $0<H\le1/2$ and $d=\kappa_2>0$, the volatility in \eqref{eq:model} satisfies $\sup_{t\ge0}\E[\sigma_t^2]<\infty$ for every $\vartheta$. More precisely, write $\alpha=H+1/2$, $C_\lambda=c_0+(c_1+\lambda)^2/(4d)$ and $\gamma_\lambda=\vartheta^2 C(\alpha)\lambda^{-2H/\alpha}$, with $C(\alpha)$ from \eqref{eq:gamma}. For every $\lambda>0$ with $\gamma_\lambda<1$,
\begin{equation}\label{eq:pointwise_bound}
\sup_{t\ge0}\E[\sigma_t^2]\le\frac{\max\{\sigma_0,C_\lambda/\lambda\}^2}{1-\gamma_\lambda}.
\end{equation}
\end{corollary}
The bound \eqref{eq:pointwise_bound} depends on volatility of volatility through $\gamma_\lambda$. The time-average bound \eqref{eq:cesaro_limit} is independent of both $\vartheta$ and $H$.

\subsection{Expected integrated variance and VIX}

The expected integrated variance follows from Proposition~\ref{prop:second} by integration,
\begin{equation}\label{eq:EI}
\E[I_T]=\int_0^T v(t)\,\D t,\qquad v(t)=\E[\sigma_t^2] .
\end{equation}
For a continuously sampled annualised variance swap the fair strike is $T^{-1}\E[I_T]$, and \eqref{eq:EI} itself is total expected variance. We express VIX in decimal volatility units as the square root of the expected average variance over the next $\Delta=30/365$ years. The quoted index in volatility points is $100\VIX$. A VIX future requires an expectation of this square root \citep{Sepp2008}.

\begin{proposition}[VIX]\label{prop:vix}
Let $\kappa_2=0$. Then
\begin{equation}\label{eq:vix}
\begin{split}
\VIX_T^2&=\frac1\Delta\E\big[I_{T+\Delta}-I_T\mid\F_T\big]=\frac1\Delta\int_T^{T+\Delta}\varpi_T(r)\,m_T(r)^2\,\D r,\\
\varpi_T(r)&=1+\int_r^{T+\Delta}\widetilde R(s-r)\,\D s ,
\end{split}
\end{equation}

where $\widetilde R$ is the positive resolvent of $\vartheta^2G^2$ defined in Proposition~\ref{prop:second}, $\widetilde R=\vartheta^2G^2+\vartheta^2G^2*\widetilde R$, and $m_T$ is the conditional mean curve \eqref{eq:cond_mean}. Under the Markovian kernel \eqref{eq:kernel_N}, $m_T$ is affine in the factor states $u_i=\sigma^{(i)}_T-\sigma_{0,i}$, and
\begin{equation}\label{eq:vix_quadratic}
\VIX_T^2=q_0+q_1^\top u+u^\top Q_2\,u
\end{equation}
with deterministic coefficients obtained by substituting the affine conditional mean in \eqref{eq:vix}. The quadratic coefficient is positive semidefinite since $\varpi_T$ is non-negative.
\end{proposition}

VIX futures require a further conditional expectation of the square root. For the finite kernel the quadratic form \eqref{eq:vix_quadratic} permits that expectation to be evaluated without nested simulation.
\section{Martingale property under positive correlation}\label{sec:measure}

\subsection{Market prices of risk proportional to volatility}

The robustness criterion of \citet{SeppRakhmonov2026} asks whether the volatility dynamics retain their form under a change of measure. For the Volterra model, the changes to the pricing and share measures each shift only the quadratic drift coefficient.

Let the physical dynamics be $\D S_t/S_t=\mu_t\,\D t+\sigma_t\,\D W^{(0)}_t$, with volatility \eqref{eq:model} and physical parameters $(\kappa_1,\kappa_2,\theta,\beta,\varepsilon)$. Following \citet{SeppRakhmonov2023}, take market prices of risk $\lambda^{(j)}_t=\bar\lambda_j\sigma_t$, $j=0,1$, and consider the density
\begin{equation}\label{eq:density}
\left.\frac{\D\Q}{\D\P}\right|_{\F_t}=\mathcal{E}\Big(-\int_0^\cdot\bar\lambda_0\sigma_u\,\D W^{(0)}_u-\int_0^\cdot\bar\lambda_1\sigma_u\,\D W^{(1)}_u\Big)_t .
\end{equation}
When this stochastic exponential is a true martingale, Girsanov's theorem gives
$\D S_t/S_t=(\mu_t-\bar\lambda_0\sigma_t^2)\D t+\sigma_t\,\D W^{\Q,(0)}_t$.
The zero-rate local-martingale price dynamics therefore require $\mu_t=\bar\lambda_0\sigma_t^2$, and a general physical drift does not admit this particular constant risk-premium specification. The true-martingale property of the price is addressed below.

\begin{proposition}[Preservation of the drift polynomial]\label{prop:closure}
If \eqref{eq:density} is a true martingale, then under $\Q$ the volatility has the same Volterra kernel and diffusion coefficients, with drift
\begin{equation}\label{eq:drift_Q}
\kappa_1\theta-(\kappa_1-\kappa_2\theta)\sigma-\kappa_2^{\Q}\sigma^2,\qquad \kappa_2^{\Q}=\kappa_2+\beta\bar\lambda_0+\varepsilon\bar\lambda_1 .
\end{equation}
Thus the intercept $c_0=\kappa_1\theta$ and linear coefficient $c_1=\kappa_2\theta-\kappa_1$ are invariant. Preservation of the polynomial form does not by itself preserve the restrictions $\kappa_1,\kappa_2\ge0$, $\theta>0$.
\end{proposition}

The proof substitutes $\D W^{(j)}=\D W^{\Q,(j)}-\bar\lambda_j\sigma\,\D t$ in the kernel integral. In particular, setting $d=\kappa_2^{\Q}$ as in \eqref{eq:sve_general}, the volatility remains in the admissible mean-reverting family whenever $c_0>0$ and $d>0$, with
\begin{equation*}
\theta^{\Q}=\frac{c_1+\sqrt{c_1^2+4dc_0}}{2d},\qquad
\kappa_1^{\Q}=\frac{\sqrt{c_1^2+4dc_0}-c_1}{2},\qquad
\kappa_2^{\Q}=d.
\end{equation*}
These parameters satisfy $(\kappa_1^{\Q}+d\sigma)(\theta^{\Q}-\sigma)=c_0+c_1\sigma-d\sigma^2$.
For $c_0>0$ and $d=0$, admissibility instead requires $c_1<0$, with $\kappa_1^{\Q}=-c_1$ and $\theta^{\Q}=c_0/(-c_1)$. If $c_0=0$, admissibility with $d>0$ requires $c_1>0$, giving $\kappa_1^{\Q}=0$ and $\theta^{\Q}=c_1/d$. With $d=0$, only $c_1=0$ is admissible and $\theta^{\Q}>0$ can be chosen arbitrarily. Negative $d$ is outside the stated family. These qualifications are independent of the true-martingale requirement on the density.

The order-flow interpretation in Appendix~\ref{sec:micro} concerns the physical drift.

\begin{corollary}[Density of the measure change]\label{cor:density}
If $\kappa_2^{\Q}\ge0$, the stochastic exponential \eqref{eq:density} is a true $\P$-martingale on $[0,T]$ for every $T>0$.
\end{corollary}

The proof in Appendix~\ref{app:proofs} repeats the localisation argument of Theorem~\ref{thm:martingale} with the density in place of the price, and the tilted equation is the volatility equation with drift \eqref{eq:drift_Q}. A negative $\kappa_2^{\Q}$ is not covered.

\subsection{Martingale property of the price}

The parameters in this subsection are those of \eqref{eq:model} under $\Q$, so that $\kappa_2$ stands for $\kappa_2^{\Q}$ of Proposition~\ref{prop:closure}. The standard localisation argument reduces the martingale property of the price to non-explosion of a tilted volatility equation. \citet[Lemma~3.1]{GuyonJeanninJourdain2026} give this reduction in their Volterra framework. In the direct representation of our model, $X=\sigma$, $F_1=F_2=\mathrm{id}$, $\Gamma_1(x)=\vartheta x$, $\Gamma_2(x)=c_0+c_1x-\kappa_2x^2$ and $K_1=K_2=K$.

For $\kappa_2=0$ their growth conditions hold. Their Theorem~1(A) then covers $\beta=0$, and their Corollary~2.6(i) covers $\beta<0$, because the volatility is non-negative.
For $\kappa_2>0$, their Assumption~1.2 fails: the assumption bounds the product of the growth exponents of $\Gamma_2$ and $F_2$ by one, and the product is two for the quadratic drift. Their Theorem~1(B) therefore does not directly establish our claim under this representation. Theorem~\ref{thm:wellposed} supplies the non-explosion step.

\begin{theorem}[Martingale property]\label{thm:martingale}
Let $K$ be the fractional kernel \eqref{eq:kernel} with $H\in(0,1/2]$ or a Markovian kernel \eqref{eq:kernel_N}. If $\kappa_2\ge\beta$, the price $S$ in \eqref{eq:model} is a true $\Q$-martingale on $[0,T]$ for every $T>0$.
\end{theorem}

Under a stopped share measure the drift gains $+\beta\sigma^2$. If $\kappa_2\ge\beta$, Theorem~\ref{thm:wellposed} excludes explosion of the tilted equation. The converse is known for the constant kernel \citep[Theorem~3.7]{SeppRakhmonov2023} and, under linear drift, for finite Markovian kernels \citep[Corollary~2.6(ii)]{GuyonJeanninJourdain2026}. The following result covers positive finite mixtures under quadratic drift as well.

\begin{proposition}[Converse for finite exponential mixtures]\label{prop:finite_converse}
Let $K=K^N$ in \eqref{eq:kernel_N} with positive weights, non-negative nodes and constant initial input $\sigma_0>0$. If $\beta>\kappa_2$, then $\E_\Q[S_T]<S_0$ for every $T>0$. Thus the price is a true martingale on every finite horizon if and only if $\kappa_2\ge\beta$.
\end{proposition}
The proof in Appendix~\ref{app:proofs} uses $K(0)<\infty$. For a singular kernel the converse remains open.

\begin{conjecture}\label{conj:supermartingale}
For the fractional kernel \eqref{eq:kernel} with $0<H<1/2$, if $\beta>\kappa_2$, then $\E_\Q[S_T]<S_0$ for every $T>0$.
\end{conjecture}

The raw conditional forward in Table~\ref{tab:martingale} has expectation one on every grid (Lemma~\ref{lem:discrete_martingale}). For $\kappa_2=0$ and $0.3$ the continuous three-factor price is a strict local martingale (Proposition~\ref{prop:finite_converse}). At 5,760 steps the two sample means are $0.93971$ and $0.98473$, with batch standard errors $0.00068$ and $0.00155$. The three means for $\kappa_2\ge\beta$ lie within $1.7$ batch standard errors of one.

Quadratic damping therefore serves two purposes. Any $\kappa_2>0$ bounds the fractional second moment, whereas the share-measure argument compares $\kappa_2$ with $\beta$ to determine whether the price is a martingale.

\section{Numerical illustrations}\label{sec:illustrations}

Figure~\ref{fig:motivation} shows the Bitcoin and Ether smiles at fixed $H=0.30$ and $H=1/2$. Each fit uses one parameter vector across three maturities and all 45 Bitcoin or 39 Ether quotes in the selected snapshot. Figures~\ref{fig:moments} and~\ref{fig:martingale} illustrate the second-moment threshold and the effect of quadratic damping.

The six risk-neutral coefficients are estimated by least squares to premium midpoints, with each error divided by its full bid--ask spread and each maturity given equal weight. We constrain $\kappa_2\ge\beta$ with a positive numerical margin. Table~\ref{tab:calibration} collects the four parameter vectors, fit errors and numerical diagnostics.

For $H=1/2$ we solve the backward equation in volatility for each Fourier variable and invert the characteristic function \citep{SeppRakhmonov2023}. The largest price change across solver checks is $0.0012$ bid--ask spreads. For $H=0.30$ we apply the bounded estimator \eqref{eq:half_tilted} on the 97-factor kernel using the classical model as a multilevel control (Appendix~\ref{sec:calibration_design}). The martingale criterion applies directly to that kernel.

The four fitted spot--volatility correlations range from $0.096$ to $0.223$. The rough and classical fits place respectively 27 and 28 of 45 Bitcoin prices, and 33 and 33 of 39 Ether prices, inside the bid--ask bands. The rough fit has a $9.4\%$ smaller weighted RMS error for Bitcoin and a $13.3\%$ larger error for Ether. Its largest pointwise 95\% price half-width is $0.04932$ spreads for Bitcoin and $0.01977$ for Ether. Both Bitcoin fits reach the imposed margin $\kappa_2-\beta=0.01$. In the rough Ether fit $\kappa_1$ is close to its lower bound. The alternative start failed the positive-eigenvalue restriction, and the table reports the best relaxed-bounds iterate at the 900-second time limit. The two snapshots illustrate the constraint at fixed $H$ and provide little information about Ether's long-run level. With $\kappa_2$ set to zero, the fractional-kernel values in Table~\ref{tab:calibration} place Bitcoin below the stability threshold and Ether above it. The corresponding 97-factor values at $H=0.30$ are $0.751$ and $8.509$.

In Figure~\ref{fig:martingale}(b), $\vartheta=1.2$ exceeds the linear three-factor threshold of $1.17$. The exact linear-drift average $\E[I_T]/(T\theta^2)$ rises from $32$ at one year to $2{,}540$ at five years. At five years Proposition~\ref{prop:cesaro} bounds the average by $19.9$ for $\kappa_2=0.3$ and $5.0$ for $\kappa_2=1.2$. The estimates across the 24 quadratic-drift cells range from $1.7$ to $2.8$. For linear drift the exact five-year grid expectation exceeds its continuous-time value by $43\%$ at 360 steps per year and $9\%$ at 720. On the finer grid the sample mean is $8.9$ against an exact $2{,}774$. The largest $0.1\%$ of paths contribute $40\%$ to $62\%$ of the linear-drift sample averages, illustrating the difficulty of estimating these moments by simulation.

\section{Conclusions and open problems}\label{sec:conclusion}

Quadratic damping permits positive spot--volatility correlation with a true price martingale. For the finite exponential mixtures used in simulation the boundary is exactly $\kappa_2=\beta$. For the singular fractional kernel we prove sufficiency, and necessity remains open. Positive quadratic damping also bounds the second moment uniformly in time. For expected integrated variance per unit time we obtain the drift-only long-run bound \eqref{eq:cesaro_limit}. The linear model has a different stability threshold, $\gamma=1$, and the simulation must use the value of $\gamma$ for its own kernel.

The trading mechanism gives these coefficients an economic interpretation. Signed volatility impacts at a rate proportional to variance generate multiplicative noise, and a small state-dependent bias generates quadratic damping. Proposition~\ref{prop:markov} proves the stopped diffusion limit for a constant kernel. The singular order-flow limit, strict positivity of the fractional volatility and its stationary law under quadratic drift remain open.
\section*{Funding}
The authors received no funding for this work.

\section*{Disclosure statement}
The authors declare that they have no competing interests.

\section*{Use of generative AI}
The authors wrote the first drafts and developed the mathematical arguments. They used OpenAI Codex (GPT-6) and Anthropic's Claude Fable 5.2 to edit the text, check the arguments, assist with the replication code and prepare the submission files. The replication scripts produced all three figures. The authors made the methodological decisions and verified the manuscript's originality and accuracy, including the references. They reviewed both tools' terms of use, confirmed their suitability for publication and take responsibility for the entire work.

\section*{Data and code availability}
The option panels use Deribit quotes for Bitcoin on 30 June 2023 and Ether on 26 October 2021, archived on 25 October 2026. The figure scripts and the simulation code are in the StochVolModels repository (\url{https://github.com/ArturSepp/StochVolModels}). Enquiries about access to the selected quotes may be addressed to the corresponding author.

\bibliographystyle{rQUF}
\bibliography{rough_volterra_lognormal_sv_qf}
\appendices

\section{Proofs}\label{app:proofs}

\begin{proof}[Proof of Lemma~\ref{lem:factors}] Write $\D Z_s=(\kappa_1+\kappa_2\sigma^N_s)(\theta-\sigma^N_s)\,\D s+\vartheta\sigma^N_s\,\D W^\sigma_s$ and $U^{i}_t=\int_0^te^{-x_i(t-s)}\D Z_s$. Then $\sigma^N_t=\sigma_0+\int_0^tK^N(t-s)\,\D Z_s=\sigma_0+\sum_iw_iU^i_t$, and It\^o's formula gives $\D U^i_t=-x_iU^i_t\,\D t+\D Z_t$. The processes $\sigma^{(i)}=\sigma_{0,i}+U^i$ solve \eqref{eq:factors}, and $\sum_iw_i\sigma^{(i)}=\sigma_0+\sum_iw_iU^i=\sigma^N$. For $s\ge t$, $\int_0^te^{-x_i(s-r)}\D Z_r=e^{-x_i(s-t)}U^i_t$, which gives \eqref{eq:theta_N}. \end{proof}

\begin{proof}[Proof of Theorem~\ref{thm:wellposed}] Write $b(x)=c_0+c_1x-dx^2$, and for $n\ge1$ define the truncated drift
\begin{equation*}
b_n(x)=b\big(\min\{\max\{x,-n\},n\}\big).
\end{equation*}
Both kernels satisfy the regularity condition (2.5) of \citet{AbiJaberLarssonPulido2019}. The fractional kernel satisfies it with exponent $2H$ by their Example~2.3(ii), and $K^N$ satisfies it with exponent one because it is bounded and Lipschitz. Both kernels are completely monotone and therefore satisfy their condition (3.4) by their Example~3.7. Constants $C$ below depend on $p$, $T$, $K$ and the coefficients, and they do not depend on $n$.

\emph{Step 1: truncation and non-negativity.} The function $b_n$ is bounded and Lipschitz, and $x\mapsto\vartheta x$ is Lipschitz. Theorem~3.3 of \citet{AbiJaberLarssonPulido2019} gives a unique continuous strong solution $\sigma^n$ of \eqref{eq:sve_general} with $b_n$ in place of the drift $b$. The coefficients are continuous with linear growth, $b_n(0)=c_0\ge0$, and the diffusion coefficient vanishes at zero. Their Theorem~3.6 therefore gives a non-negative continuous weak solution, and pathwise uniqueness identifies its law with the law of $\sigma^n$. Thus $\sigma^n_t\ge0$ for all $t$, almost surely.

\emph{Step 2: one-sided bound.} For $x\ge0$ we have $b_n(x)\le c_0+c_1^+x$, where $c_1^+=\max(c_1,0)$. Since $K\ge0$ and $\sigma^n\ge0$,
\begin{equation}\label{eq:one_sided}
0\le\sigma^n_t\le\sigma_0+c_0\int_0^tK(u)\,\D u+c_1^+\int_0^tK(t-s)\sigma^n_s\,\D s+N^n_t,\qquad N^n_t=\vartheta\int_0^tK(t-s)\sigma^n_s\,\D B_s .
\end{equation}
The quadratic term has dropped out of the upper bound.

\emph{Step 3: moments uniform in $n$.} Fix $p\ge2$ and $T>0$, and let $f_n(t)=\E[(\sigma^n_t)^p]$, which is bounded on $[0,T]$ by Lemma~3.1 of \citet{AbiJaberLarssonPulido2019}. Jensen's inequality gives $\big(\int_0^tK(t-s)\sigma^n_s\,\D s\big)^p\le\big(\int_0^TK(u)\,\D u\big)^{p-1}\int_0^tK(t-s)(\sigma^n_s)^p\,\D s$. For fixed $t$, $N^n_t$ is the terminal value of the martingale $u\mapsto\vartheta\int_0^uK(t-s)\sigma^n_s\,\D B_s$ on $[0,t]$. The Burkholder--Davis--Gundy and Jensen inequalities give $\E|N^n_t|^p\le C\int_0^tK(t-s)^2f_n(s)\,\D s$. Taking the $p$-th power and expectations in \eqref{eq:one_sided} yields
\begin{equation*}
f_n(t)\le C+C\int_0^t\big(K(t-s)+K(t-s)^2\big)f_n(s)\,\D s,\qquad t\le T .
\end{equation*}
Choose $\varrho>0$ with $C\int_0^Te^{-\varrho u}\big(K(u)+K(u)^2\big)\D u\le1/2$, which is possible because $K+K^2$ is integrable on $[0,T]$. Multiplying the inequality by $e^{-\varrho t}$ and taking the supremum over $t\le T$ gives $\sup_{t\le T}e^{-\varrho t}f_n(t)\le2C$, so that $\sup_n\sup_{t\le T}f_n(t)\le2Ce^{\varrho T}$.

\emph{Step 4: supremum bound.} On $x\ge0$, $|b_n(x)|\le c_0+|c_1|x+dx^2$, so Step~3 with $2p$ in place of $p$ bounds $\sup_{t\le T}\E|b_n(\sigma^n_t)|^p$ uniformly in $n$. Lemma~2.4 of \citet{AbiJaberLarssonPulido2019} then gives $\E\big[\sup_{t\le T}(\sigma^n_t)^p\big]\le C$ for $p>\max\{2,1/H\}$, and for $p>2$ in the case of $K^N$. With $\tau_n=\inf\{t\ge0:\sigma^n_t\ge n\}$, Markov's inequality gives
\begin{equation}\label{eq:tau_n}
\P(\tau_n\le T)\le Cn^{-p} .
\end{equation}

\emph{Step 5: local uniqueness and removal of the truncation.} Let $X$ and $Y$ be continuous adapted processes and let $\tau$ be the first time that $|X|$ or $|Y|$ reaches $n$. Assume that, up to $\tau$, both satisfy \eqref{eq:sve_general} with the same Brownian motion and with drifts that coincide with $b$ on $[-n,n]$. On $\{t\le\tau\}$ both drift integrands equal $b$ evaluated in $[-n,n]$, where $b$ is Lipschitz, and the indicator $\one_{\{s\le\tau\}}$ can be inserted in both integrals. The function $\Delta(t)=\E\big[|X_t-Y_t|^2\one_{\{t\le\tau\}}\big]$ therefore satisfies $\Delta(t)\le C_n\int_0^t\big(K(t-s)+K(t-s)^2\big)\Delta(s)\,\D s$, and the argument of Step~3 gives $\Delta\equiv0$, that is, $X=Y$ on $[0,\tau]$. In particular $\sigma^m=\sigma^n$ on $[0,\tau_n]$ for $m\ge n$, the times $\tau_n$ increase, and $\tau_n\to\infty$ almost surely by \eqref{eq:tau_n}. The limit $\sigma=\lim_n\sigma^n$ is continuous and non-negative. The limit solves \eqref{eq:sve_general} on $[0,\infty)$, because $b_n(\sigma^n)=b(\sigma)$ on $[0,\tau_n]$ and stochastic integrals of integrands that agree on an event agree on that event. The same local argument gives pathwise uniqueness among continuous solutions. The moment bound follows from Step~4 and Fatou's lemma, first for large $p$ and then for every $p\ge1$ by Jensen's inequality. \end{proof}

\begin{proof}[Proof of Proposition~\ref{prop:mean}] For $\kappa_2=0$ the coefficients are globally Lipschitz with linear growth. Lemma~3.1 of \citet{AbiJaberLarssonPulido2019} gives $\sup_{s\le T}\E[\sigma_s^2]<\infty$ on every finite horizon, and their Example~2.3(ii) verifies the kernel regularity. The stochastic integral in \eqref{eq:flow} is square integrable since $K\in L^2_{\rm loc}$, and has zero conditional mean. Therefore $m_t(s)=\Theta^t_s+\kappa_1\int_t^sK(s-r)(\theta-m_t(r))\,\D r$. Writing $\phi=m_t-\theta$ gives $\phi=\psi_t-\kappa_1K*\phi$ on $[t,\infty)$, so \eqref{eq:resolvent} yields $\phi=\psi_t-R*\psi_t$. At $t=0$, $\psi_0\equiv\sigma_0-\theta$, giving the unconditional mean. For the fractional kernel, $\widehat R(z)=\kappa_1/(z^\alpha+\kappa_1)$, which inverts to \eqref{eq:resolvent_frac}. When $\kappa_1=0$, $R=0$. \end{proof}

\begin{proof}[Proof of Proposition~\ref{prop:second}] Subtracting the equation for $m_t$ from \eqref{eq:flow}, the centred process $\chi_s=\sigma_s-m_t(s)$ solves $\chi=-\kappa_1K*\chi+K*\D M$ on $[t,\infty)$, with $\D M_r=\vartheta\sigma_r\,\D W^\sigma_r$. The process $G*\D M$ solves the same linear equation, because $G=K-\kappa_1K*G$, by \eqref{eq:resolvent} when $\kappa_1>0$ and directly from $G=K$ when $\kappa_1=0$, together with the stochastic Fubini theorem. Uniqueness for the linear equation gives $\chi_s=\int_t^sG(s-r)\vartheta\sigma_r\,\D W^\sigma_r$. It\^o's isometry yields $\E[\chi_s^2\mid\F_t]=\vartheta^2\int_t^sG(s-r)^2v_t(r)\,\D r$, and $v_t(s)=m_t(s)^2+\E[\chi_s^2\mid\F_t]$ because $\E[\chi_s\mid\F_t]=0$. The a priori bound in the proof of Proposition~\ref{prop:mean} makes every term finite. \end{proof}

\begin{proof}[Proof of Corollary~\ref{cor:ml}] With $\kappa_1=0$, $G=K$ and $m\equiv\sigma_0$. Since $\vartheta^2K(t)^2=a\,t^{2H-1}/\Gamma(2H)$, equation \eqref{eq:cond_second} at $t=0$ reads $v=\sigma_0^2+a\,\mathcal{I}^{2H}v$, where $\mathcal{I}^{2H}$ is the Riemann--Liouville integral of order $2H$. Its solution is $\sigma_0^2E_{2H}(a\,t^{2H})$, and integration term by term gives $\E[I_T]$. The asymptotics follow from $E_{2H}(z)\sim(2H)^{-1}\exp(z^{1/(2H)})$ as $z\to+\infty$ \citep{GorenfloKilbasMainardiRogosin2014}. \end{proof}

\begin{proof}[Proof of Proposition~\ref{prop:vix}] For continuous paths, $-\tfrac2\Delta\E[\ln(S_{T+\Delta}/S_T)\mid\F_T]=\tfrac1\Delta\E[I_{T+\Delta}-I_T\mid\F_T]=\tfrac1\Delta\int_T^{T+\Delta}v_T(s)\,\D s$. Substituting $v_T=m_T^2+\widetilde R*m_T^2$ from Proposition~\ref{prop:second} and exchanging the order of integration gives \eqref{eq:vix}. Under $K^N$, \eqref{eq:theta_N} makes $\psi_T$ affine in $u$, and \eqref{eq:cond_mean} preserves affinity. The square $m_T^2$ is then a quadratic form in $u$, and so is its integral against $\varpi_T$. \end{proof}

\begin{proof}[Proof of Corollary~\ref{cor:pointwise}] Completing the square gives $b(x)=c_0+c_1x-dx^2\le C_\lambda-\lambda x$. Let $Y$ solve the equation with this affine drift and the same kernel, Brownian motion, multiplicative diffusion and initial value. To apply the scalar comparison theorem of \citet[Theorem~3.3]{Alfonsi2025}, use $b_n(x)=\min\{b((-n)\vee(x\wedge n)),C_\lambda-\lambda x\}$. Both comparison drifts are globally Lipschitz and $b_n=b$ on $[-n,n]$. The fractional kernel is completely monotone and satisfies the integrability and increment condition of that theorem for $H>0$. The constant kernel admits the usual scalar diffusion comparison. Local uniqueness up to the first exit from $[-n,n]$ and non-explosion in Theorem~\ref{thm:wellposed} yield $0\le\sigma_t\le Y_t$ for all $t$. The mean of $Y$ lies between $\sigma_0$ and $C_\lambda/\lambda$. On every finite horizon its second-moment renewal equation \eqref{eq:cond_second} bounds $\sup_{t\le T}\E[Y_t^2]$ by $\max\{\sigma_0,C_\lambda/\lambda\}^2+\gamma_\lambda\sup_{t\le T}\E[Y_t^2]$. Rearrangement gives \eqref{eq:pointwise_bound} uniformly in $T$. Since $2H/\alpha>0$, a sufficiently large $\lambda$ makes $\gamma_\lambda<1$. \end{proof}

\begin{proof}[Proof of Proposition~\ref{prop:finite_converse}] Write $\widehat\sigma$ for the auxiliary volatility with share-measure drift $c_0+c_1\widehat\sigma+(\beta-\kappa_2)\widehat\sigma^2$, up to its explosion time $\zeta$. Its finite-factor lift has locally Lipschitz coefficients. Let $k=K(0)>0$, let $L$ be the first-kind resolvent, and put $M=K'*L$. The stopped decomposition of \citet{BondiPulido2026} (Lemma~1 and Remark~2 in the preprint) gives, up to exit from a bounded interval,
\[
\D\widehat\sigma_t=(a\widehat\sigma_t+q\widehat\sigma_t^2+A_t)\,\D t+v\widehat\sigma_t\,\D B_t,\qquad
A_t=kc_0+\int_{(0,t]}\widehat\sigma_{t-s}\,\D M(s)-\sigma_0M(t),
\]
where $a=kc_1+K'(0)/k$, $q=k(\beta-\kappa_2)>0$ and $v=k\vartheta>0$. Complete monotonicity gives $M\le0$ and $\D M\ge0$, so $A_t\ge0$ while $\widehat\sigma$ stays positive. With $E_t=\exp((a-v^2/2)t+vB_t)$, the process $Y=\widehat\sigma/E$ starts at $\sigma_0$ and satisfies $Y'\ge qE Y^2$. Hence $\widehat\sigma$ stays positive until explosion and $1/Y_t\le1/\sigma_0-q\int_0^tE_s\,\D s$. Brownian motion has positive probability to stay in a uniform tube about a sufficiently steep line on $[0,T]$. On that event $q\int_0^T E_s\,\D s>1/\sigma_0$, so $\zeta\le T$. A bounded $\widehat\sigma$ gives bounded drift and diffusion integrands in the variation-of-constants formula for every factor on a finite horizon. Thus the lift cannot explode while $\widehat\sigma$ stays bounded. By local uniqueness the law up to the first upper level $n$ agrees with the stopped share-measure law used in the proof of Theorem~\ref{thm:martingale}. Passing to increasing levels in that density identity gives $\E_\Q[S_T]/S_0=\Pr_{\mathrm{aux}}(\zeta>T)<1$, where $\Pr_{\mathrm{aux}}$ is the law of the auxiliary lift. \end{proof}

\begin{proof}[Proof of Theorem~\ref{thm:martingale}]
The stochastic exponential $S/S_0$ is a positive local martingale and therefore a supermartingale. By Theorem~\ref{thm:wellposed} the volatility is non-negative and does not explode. For $n>\sigma_0$, define $\tau_n=\inf\{t\ge0:\sigma_t\ge n\}$, with $\inf\varnothing=\infty$. The stopped price is a martingale by Novikov's criterion, because $\int_0^{T\wedge\tau_n}\sigma_s^2\D s\le n^2T$. Define
\begin{equation*}
\left.\frac{\D\widehat\Q^{(n)}}{\D\Q}\right|_{\F_T}=\frac{S_{T\wedge\tau_n}}{S_0}.
\end{equation*}
Then $\widehat W^{(0,n)}_t=W^{(0)}_t-\int_0^{t\wedge\tau_n}\sigma_u\D u$ and $W^{(1)}$ form a two-dimensional Brownian motion under $\widehat\Q^{(n)}$, and $\widehat B=(\beta\widehat W^{(0,n)}+\varepsilon W^{(1)})/\vartheta$ is a Brownian motion. Up to $\tau_n$, the volatility satisfies \eqref{eq:sve_general} with respect to $\widehat B$, with drift
\begin{equation*}
(\kappa_1+\kappa_2\sigma)(\theta-\sigma)+\beta\sigma^2
=\kappa_1\theta-(\kappa_1-\kappa_2\theta)\sigma-(\kappa_2-\beta)\sigma^2.
\end{equation*}
Let $\widehat\sigma$ be the global solution of \eqref{eq:sve_general} driven by $\widehat B$ with $c_0=\kappa_1\theta$, $c_1=\kappa_2\theta-\kappa_1$ and $d=\kappa_2-\beta\ge0$, which exists by Theorem~\ref{thm:wellposed}. Let $\widehat\tau_n$ be the first time that $\widehat\sigma$ reaches $n$. Step~5 of the proof of Theorem~\ref{thm:wellposed} gives $\sigma=\widehat\sigma$ on $[0,\tau_n\wedge\widehat\tau_n]$, and therefore $\tau_n=\widehat\tau_n$. The law of $\widehat\sigma$ does not depend on $n$, so the moment bound of Theorem~\ref{thm:wellposed} and Markov's inequality give $\widehat\Q^{(n)}(\tau_n\le T)\le Cn^{-p}$ with $C$ independent of $n$. Thus
\begin{equation*}
\frac{\E_{\Q}[S_T\one_{\{\tau_n>T\}}]}{S_0}
=\widehat\Q^{(n)}(\tau_n>T)
\ge1-Cn^{-p}\longrightarrow1.
\end{equation*}
Since $\one_{\{\tau_n>T\}}\uparrow1$ almost surely, monotone convergence yields $\E_{\Q}[S_T]=S_0$. A non-negative supermartingale whose expectation is unchanged over $[0,T]$ is a martingale, proving the claim. \end{proof}

\begin{proof}[Proof of Corollary~\ref{cor:density}] Denote the stochastic exponential in \eqref{eq:density} by $D$. The process $D$ is a positive $\P$-local martingale, and with $\tau_n$ as in the previous proof the stopped process $D_{\cdot\wedge\tau_n}$ is a martingale by Novikov's criterion. Under the measure with density $D_{T\wedge\tau_n}$, the substitution that proves Proposition~\ref{prop:closure} shows that, up to $\tau_n$, the volatility satisfies \eqref{eq:sve_general} with $c_0=\kappa_1\theta$, $c_1=\kappa_2\theta-\kappa_1$ and $d=\kappa_2^{\Q}$. For $\kappa_2^{\Q}\ge0$ the argument of the previous proof gives $\E_{\P}[D_T\one_{\{\tau_n>T\}}]\ge1-Cn^{-p}$, and therefore $\E_{\P}[D_T]=1$. \end{proof}

\begin{proof}[Proof of Proposition~\ref{prop:stationarity}] The mean $m(t)=\theta+(\sigma_0-\theta)E_\alpha(-\kappa_1t^\alpha)$ lies between $\sigma_0$ and $\theta$, so $m^2$ is bounded above and below by a positive constant, and $m^2\to\theta^2$. The kernel $h=\vartheta^2G^2$ is non-negative and integrable: $G(t)\sim t^{\alpha-1}/\Gamma(\alpha)$ at zero with $\alpha>1/2$, and $G$ decays as $O(t^{-1-\alpha})$ for $\alpha<1$ and exponentially at $\alpha=1$. Its positive resolvent has the locally convergent convolution series $\widetilde R=\sum_{j\ge1}h^{*j}$. Tonelli's theorem gives
\begin{equation*}
\int_0^\infty\widetilde R(s)\,\D s
=\sum_{j\ge1}\Big(\int_0^\infty h(s)\,\D s\Big)^j
=\sum_{j\ge1}\gamma^j.
\end{equation*}
If $\gamma<1$, Proposition~\ref{prop:second} gives $\sup v\le\sup m^2/(1-\gamma)$, and dominated convergence in $\widetilde R*m^2$ gives $v(t)\to\theta^2/(1-\gamma)$. If $\gamma\ge1$, then with $\underline m=\min(\sigma_0,\theta)>0$,
$v(t)\ge\underline m^2(1+\int_0^t\widetilde R(s)\,\D s)\to\infty$.

This direction needs only $\theta\ne0$. If $m(s)^2\ge\theta^2/4$ for $s\ge s_0$, then $v(t)\ge(\theta^2/4)\int_0^{t-s_0}\widetilde R(s)\,\D s\to\infty$.
For the norm, $\widehat G(z)=1/(z^\alpha+\kappa_1)$ and Parseval's identity imply
\begin{equation*}
\int_0^\infty G^2=\frac1{2\pi}\int_{-\infty}^\infty
\frac{\D\omega}{|\omega|^{2\alpha}+2\kappa_1|\omega|^\alpha\cos(\pi\alpha/2)+\kappa_1^2}.
\end{equation*}
The substitution $\omega=\kappa_1^{1/\alpha}u$ gives the factor $\kappa_1^{1/\alpha-2}$. For $\alpha<1$, substituting $y=u^\alpha$ in the remaining integral and integrating the two complex-conjugate linear factors gives
\begin{equation*}
C(\alpha)=-\frac{\cot(\pi\alpha/2)}{\alpha\sin(\pi/\alpha)}.
\end{equation*}
At $\alpha=1$, direct integration gives $C(1)=1/2$, also the limit of this expression. \end{proof}

\begin{proof}[Proof of Proposition~\ref{prop:growth}] Parseval's identity gives the second expression in \eqref{eq:h_hat}. We apply it to $u\mapsto e^{-\lambda u/2}G(u)$, whose Laplace transform on the imaginary axis is
\begin{equation*}
\widehat G(\lambda/2+\mathrm{i}\omega)=\big((\lambda/2+\mathrm{i}\omega)^\alpha+\kappa_1\big)^{-1}.
\end{equation*}
The function $\widehat h$ is continuous and strictly decreasing, and $\widehat h(0+)=\gamma$ by monotone convergence. For $\kappa_1>0$, dominated convergence gives $\widehat h(\lambda)\to0$ as $\lambda\to\infty$, and for $\kappa_1=0$ the transform is $a\lambda^{-2H}$ with $a$ from \eqref{eq:ml_second}. For $\gamma>1$ the root $\lambda^*$ therefore exists and is unique. Equation \eqref{eq:cond_second} at $t=0$ reads $v=m^2+h*v$. Multiplying by $e^{-\lambda^*t}$ gives the renewal equation $v_\ast=z+h_\ast*v_\ast$ for $v_\ast(t)=e^{-\lambda^*t}v(t)$, with $z(t)=e^{-\lambda^*t}m(t)^2$ and $h_\ast(t)=e^{-\lambda^*t}h(t)$. The function $h_\ast$ is a probability density with finite mean $\int_0^\infty u\,h_\ast(u)\,\D u=|\widehat h'(\lambda^*)|$. The function $z$ is continuous and bounded by a multiple of $e^{-\lambda^*t}$, and is therefore directly Riemann integrable. The key renewal theorem \citep[Chapter~XI]{Feller1971} gives $v_\ast(t)\to|\widehat h'(\lambda^*)|^{-1}\int_0^\infty z(s)\,\D s$, which is \eqref{eq:growth}. For $\sigma_0=\theta$ we have $m\equiv\theta$ and $\int_0^\infty z(s)\,\D s=\theta^2/\lambda^*$.

For $\gamma=1$ the kernel $h$ itself is a probability density. Its mean $\mu_h$ is finite, because $u\,G(u)^2$ is of order $u^{2\alpha-1}$ at zero and of order $u^{-1-2\alpha}$ at infinity for $\alpha<1$, and decays exponentially at $\alpha=1$. The solution is $v=m^2+\widetilde R*m^2$, where $\widetilde R=\sum_{j\ge1}h^{*j}$ is the renewal density. The renewal theorem gives $t^{-1}\int_0^t\widetilde R(s)\,\D s\to1/\mu_h$. Since $m^2$ is bounded and $m^2\to\theta^2$, splitting the convolution at a fixed distance from $t$ gives $v(t)/t\to\theta^2/\mu_h$. \end{proof}

\begin{proof}[Proof of Proposition~\ref{prop:markov}] For $K\equiv1$, set $\delta=\vartheta/\sqrt n$. The Markov chain $(\sigma^n,P^n)$ has jumps $(\delta\zeta,n^{-1/2}\eta)$ at rate $n\sigma^2$, where $\zeta$ and $\eta$ denote the marks of a trade. For $f\in C_c^\infty((0,\infty)\times\mathbb R)$ its generator is $A_nf(\sigma,p)=n\sigma^2\,\E\big[f(\sigma+\delta\zeta,p+n^{-1/2}\eta)-f(\sigma,p)\big]$. Taylor expansion and the mark moments in \eqref{eq:bias} give
\begin{equation*}
A_nf=\sqrt n\,\vartheta\sigma^2\bar\zeta_n\,\partial_\sigma f+\tfrac12\vartheta^2\sigma^2\,\partial_{\sigma\sigma}f+\tfrac12\sigma^2\,\partial_{pp}f+\rho(1-|\bar\zeta_n|)\vartheta\sigma^2\,\partial_{\sigma p}f+O(n^{-1/2}),
\end{equation*}
uniformly on compact subsets of $(0,\infty)\times\mathbb R$. Fix the domain $D$ of the proposition. On a neighbourhood of its closure the clip is inactive for all sufficiently large $n$, so the first coefficient equals $(\kappa_1+\kappa_2\sigma)(\theta-\sigma)$ and $|\bar\zeta_n|\to0$ uniformly. Thus $A_n$ converges there to the generator of \eqref{eq:markov_limit}. The stopped drift and predictable quadratic-variation rates are uniformly bounded, and the jump sizes tend to zero. These bounds give tightness of the stopped martingale problems, and every limit has continuous paths. The limiting coefficients are locally Lipschitz. On $\overline D$ the covariance matrix is uniformly positive definite because $\sigma\ge\sigma_->0$ and $1-\rho^2=\varepsilon^2/\vartheta^2>0$. Therefore the stopped diffusion law is unique, and the boundary of the rectangle is regular for exit. Localisation of the diffusion-approximation argument of \citet{EthierKurtz1986} identifies the limit as the diffusion stopped on leaving $D$, and the chain's exit overshoot tends to zero with its jump size. These steps prove the stated stopped convergence. Removing the stops requires uniform exit estimates for the chains as $\sigma_-\downarrow0$ and $\sigma_+,\bar p\uparrow\infty$, which are not supplied here. \end{proof}

\begin{proof}[Proof of Proposition~\ref{prop:classical_qd}] For $K\equiv1$ the volatility is the diffusion $\D\sigma=b(\sigma)\,\D t+\vartheta\sigma\,\D W^\sigma$ on $(0,\infty)$, where $b(\sigma)=\kappa_1\theta-(\kappa_1-\kappa_2\theta)\sigma-\kappa_2\sigma^2$. Its speed density is proportional to
\begin{equation*}
\sigma^{-2}\exp\left(\int \frac{2b(\sigma)}{\vartheta^2\sigma^2}\,\D\sigma\right),
\end{equation*}
which gives \eqref{eq:stationary_density}. Put $A=2\kappa_1\theta/\vartheta^2>0$, $B=2\kappa_2/\vartheta^2>0$ and $c=2(\kappa_1-\kappa_2\theta)/\vartheta^2$. The scale and speed densities are proportional to $\sigma^c e^{A/\sigma+B\sigma}$ and $\sigma^{-c-2}e^{-A/\sigma-B\sigma}$. The scale integrals diverge at both endpoints, excluding access from the interior, while the entrance integrals are finite, so both endpoints are entrance boundaries. The finite speed measure normalises to the stationary law. The factors $e^{-B\sigma}$ and $e^{-A/\sigma}$ make all positive and negative stationary moments finite.
For a deterministic initial state, the generator satisfies
\begin{equation*}
\mathcal L\sigma^2=2\kappa_1\theta\sigma+
(-2\kappa_1+2\kappa_2\theta+\vartheta^2)\sigma^2-2\kappa_2\sigma^3
\le C-c_2\sigma^2
\end{equation*}
for some $C,c_2>0$. The stopped It\^o estimate and localisation therefore give
$\E[\sigma_t^2]\le e^{-c_2t}\sigma_0^2+C(1-e^{-c_2t})/c_2$, proving uniform boundedness. \end{proof}

\begin{proof}[Proof of Proposition~\ref{prop:cesaro}] Let $b(x)=\kappa_1\theta+c_1x-\kappa_2x^2$ with $c_1=\kappa_2\theta-\kappa_1$, and define the semimartingale $Z_t=\int_0^tb(\sigma_s)\,\D s+\vartheta\int_0^t\sigma_s\,\D W^\sigma_s$, so that $\sigma-\sigma_0=K*\D Z$. Both kernels are completely monotone. By Example~3.7 of \citet{AbiJaberLarssonPulido2019} they satisfy their condition (3.4), which includes a non-negative resolvent of the first kind $L$. For the fractional kernel with $\alpha<1$, $L(\D s)=s^{-\alpha}\,\D s/\Gamma(1-\alpha)$ and $L([0,T])=T^{1-\alpha}/\Gamma(2-\alpha)$. At $\alpha=1$, $L=\delta_0$ and $L([0,T])=1$, which is the same expression. Their Lemma~2.6 gives $Z=L*(\sigma-\sigma_0)$. Since $\sigma\ge0$ by Theorem~\ref{thm:wellposed} and $L\ge0$,
\begin{equation}\label{eq:Z_lower}
Z_T=\int_{[0,T]}(\sigma_{T-s}-\sigma_0)\,L(\D s)\ge-\sigma_0\,L([0,T]) .
\end{equation}
The moments of Theorem~\ref{thm:wellposed} make the stochastic integral in $Z$ a martingale. With $m(t)=\E[\sigma_t]$ and $v(t)=\E[\sigma_t^2]$, taking expectations in \eqref{eq:Z_lower} gives
\begin{equation*}
\kappa_2\int_0^Tv(t)\,\D t\le\kappa_1\theta\,T+c_1\int_0^Tm(t)\,\D t+\sigma_0\,L([0,T]) .
\end{equation*}
If $c_1\le0$ we drop the middle term, because $m\ge0$. If $c_1>0$ we use $m\le\sqrt v$ and the Cauchy--Schwarz inequality, $\int_0^Tm(t)\,\D t\le\big(T\int_0^Tv(t)\,\D t\big)^{1/2}=T\sqrt{A_T}$. Division by $T$ gives \eqref{eq:cesaro}, which is a quadratic inequality in $\sqrt{A_T}$. Thus $\sqrt{A_T}$ is at most the positive root $y_T$ of $\kappa_2y^2-c_1^+y-\kappa_1\theta-\sigma_0L([0,T])/T=0$, and $y_T$ increases with the last term.

For the fractional kernel the last term tends to zero as $T\to\infty$. If $c_1\le0$, then $y_T^2\to\kappa_1\theta/\kappa_2$, and $\kappa_1/\kappa_2\ge\theta$ in this case. If $c_1>0$, then $y_T\to\big(c_1+\sqrt{c_1^2+4\kappa_2\kappa_1\theta}\big)/(2\kappa_2)=\theta$, because $c_1^2+4\kappa_2\kappa_1\theta=(\kappa_2\theta+\kappa_1)^2$, and $\theta>\kappa_1/\kappa_2$ in this case. Both cases give \eqref{eq:cesaro_limit}.

For $K^N$ put $k(t)=\int_0^tK^N(s)\,\D s$. We integrate $K^N*L\equiv1$ over $[0,T]$ and exchange the order of integration, which gives $T=\int_{[0,T]}k(T-s)\,L(\D s)$ for every $T>0$. The function $k$ is non-decreasing and $L\ge0$, so that $T\le k(T)\,L([0,T])$. The same identity at $T+a$ gives $T+a\ge k(a)\,L([0,T])$ for every $a>0$, because $k(T+a-s)\ge k(a)$ for $s\le T$. Thus
\begin{equation*}
\frac{1}{k(T)}\le\frac{L([0,T])}{T}\le\frac{1+a/T}{k(a)} .
\end{equation*}
We let $T\to\infty$ and then $a\to\infty$ and obtain $L([0,T])/T\to1/k(\infty)=\ell_N$. The limit of $y_T^2$ is the right-hand side of \eqref{eq:cesaro_limit_N}. \end{proof}

\section{Simulation scheme and validation}\label{app:simulation}

\subsection{Symmetric splitting of the factor system}\label{sec:splitting}

We approximate the factor system \eqref{eq:factors} on a uniform grid $t_n=nh$. In this appendix $h$ denotes the time step and no longer the kernel of Section~\ref{sec:stability}. In this appendix grid-time state symbols denote the numerical approximation, rather than the exact solution. Every path carries all $N$ factors, so the state at $t_n$ contains the full history of the Markovian model.

Write $c_0=\kappa_1\theta$ and $c_1=\kappa_2\theta-\kappa_1$ as in \eqref{eq:sve_general}. The Stratonovich form of \eqref{eq:factors} is
\begin{equation}\label{eq:factors_strat}
\D\sigma^{(i)}_t=\Big[-x_i\big(\sigma^{(i)}_t-\sigma_{0,i}\big)+c_0+\tilde c_1\sigma^N_t-\kappa_2\big(\sigma^N_t\big)^2\Big]\D t+\vartheta\sigma^N_t\circ\D W^\sigma_t,\qquad \tilde c_1=c_1-\tfrac12\bar w\vartheta^2 .
\end{equation}
The correction $-\tfrac12\bar w\vartheta^2\sigma^N$ is the same for every factor, because every factor receives the noise $\vartheta\sigma^N\D W^\sigma$ and $\sigma^N$ has the derivative $w_j$ with respect to $\sigma^{(j)}$.

We split the right-hand side of \eqref{eq:factors_strat} into a linear part, the quadratic damping and the noise. Each part has a flow in closed form, and the linear part contains the decay rates $x_i$, which are the stiff terms of the system. Let $\boldsymbol U=(U^1,\dots,U^N)^\top$ with $U^i=\sigma^{(i)}-\sigma_{0,i}$ denote the centred factors of the proof of Lemma~\ref{lem:factors}, so that $\sigma^N=\sigma_0+w^\top\boldsymbol U$, and let $\one$ denote the vector of ones. The memory map $\mathcal{M}_\tau$ solves the linear part $\D\sigma^{(i)}/\D t=-x_i(\sigma^{(i)}-\sigma_{0,i})+c_0+\tilde c_1\sigma^N$ over a time $\tau$,
\begin{equation}\label{eq:memory_map}
\mathcal{M}_\tau:\ \boldsymbol U\mapsto\boldsymbol U^\ast+e^{-\Lambda\tau}\big(\boldsymbol U-\boldsymbol U^\ast\big),\qquad \Lambda=\mathrm{diag}(x)-\tilde c_1\one w^\top,\qquad \boldsymbol U^\ast=(c_0+\tilde c_1\sigma_0)\,\Lambda^{-1}\one .
\end{equation}
The damping map $\mathcal{D}_\tau$ solves $\D\sigma^{(i)}/\D t=-\kappa_2(\sigma^N)^2$ over a time $\tau$, and the noise map $\mathcal{N}_h$ solves $\D\sigma^{(i)}=\vartheta\sigma^N\circ\D W^\sigma$ over one step. Both maps act on the weighted sum,
\begin{equation}\label{eq:scalar_maps}
\mathcal{D}_\tau:\ \sigma^N\mapsto\frac{\sigma^N}{1+\bar w\kappa_2\tau\,\sigma^N},\qquad
\mathcal{N}_h:\ \sigma^N\mapsto\sigma^N\exp\big(\bar w\vartheta\sqrt h\,Z^\sigma\big),
\end{equation}
with $Z^\sigma$ standard normal, and both add the same amount to every factor, which is the change of $\sigma^N$ divided by $\bar w$. One step of the scheme is the symmetric composition
\begin{equation}\label{eq:scheme}
\boldsymbol U_{n+1}=\mathcal{M}_{h/2}\circ\mathcal{D}_{h/2}\circ\mathcal{N}_h\circ\mathcal{D}_{h/2}\circ\mathcal{M}_{h/2}\big(\boldsymbol U_n\big),
\end{equation}
with one normal variable $Z^\sigma_n$ per step.

The memory map integrates the decay rates exactly, so the largest node does not restrict the time step. For the kernel of the calibration the largest node is $7.4\times10^6$ per year. Conjugation with $\mathrm{diag}(\sqrt{w_i})$ makes $\Lambda$ symmetric, and the implementation diagonalises the symmetric matrix once for each parameter vector and time step.\footnote{Functions \texttt{prepare\_history} and \texttt{simulate\_history} of the module \path{src/stochvolmodels/models/rough_logsv/history.py}.} The implementation requires the smallest eigenvalue of $\Lambda$ to be positive, which holds whenever $\tilde c_1<0$ and all nodes are positive. This requirement belongs to the numerical method, and Theorem~\ref{thm:wellposed} does not need it.

The damping and noise maps keep a positive $\sigma^N$ positive. The memory map is affine and does not guarantee a positive $\sigma^N$ for $N>1$. The implementation stops with an error if a volatility state is non-positive or not finite, and uses no floor, no reset of the state and no deletion of paths. All runs behind Tables~\ref{tab:martingale} and~\ref{tab:calibration} and Figures~\ref{fig:motivation} and~\ref{fig:martingale} completed without such an error.

Each of the three flows is exact, and their composition \eqref{eq:scheme} approximates the solution of \eqref{eq:factors}. \citet{NinomiyaVictoir2008} compose the flows of the Stratonovich vector fields in their weak scheme of second order, and \citet{BayerBreneis2024} use splitting schemes of this type for the Markovian approximation of rough Heston. \citet{Shinozaki2026} splits the drift into the decay of the factors and the remaining drift, and proves weak order two for vector fields with bounded derivatives. For rough Heston he also proves that the variance stays non-negative under a parameter condition. For the volatility factors, the composition \eqref{eq:scheme} has the same symmetric form with a different grouping of the drift. We keep the linear drift in the memory map, because the map then integrates the whole drift exactly when $\kappa_2=0$. His composition also contains the price, which we advance by the separate step \eqref{eq:price_step}. The quadratic drift is outside the assumptions of his theorem. We prove neither a convergence order nor non-negativity for \eqref{eq:scheme}. Section~\ref{sec:validation} measures its time-discretisation error against exact moments.

\subsection{Price step and the discrete martingale}\label{sec:price_step}

The price step holds the volatility at its value $\sigma^N_n$ at the beginning of the step,
\begin{equation}\label{eq:price_step}
\begin{split}
I_{n+1}&=I_n+\big(\sigma^N_n\big)^2h,\qquad J_{n+1}=J_n+\sigma^N_n\sqrt h\,Z^\sigma_n,\\
X_{n+1}&=X_n-\tfrac12\big(\sigma^N_n\big)^2h+\sigma^N_n\sqrt h\,\Big(\rho Z^\sigma_n+\sqrt{1-\rho^2}\,Z^\perp_n\Big),
\end{split}
\end{equation}
with $I_0=J_0=0$. Here $Z^\sigma_n$ is the normal variable of the noise map in step $n$, and $Z^\perp_n$ is a standard normal variable that is independent of all other variables. The sums $I_n$ and $J_n$ are the left-point sums of $\int(\sigma^N_u)^2\,\D u$ and $\int\sigma^N_u\,\D W^\sigma_u$. Beginning-of-step volatility is predictable for the It\^o sum $J_n$ and gives the conditional martingale identity below.

\begin{lemma}[Discrete martingale]\label{lem:discrete_martingale}
Suppose that the volatility states $\sigma^N_n$ of the scheme \eqref{eq:scheme} are finite almost surely. Then, for every step $h>0$, every $\kappa_2\ge0$ and every $\rho\in[-1,1]$, the discrete price $S_n=S_0e^{X_n-X_0}$ of \eqref{eq:price_step} is a martingale, and $\E[S_n]=S_0$ for every $n$. The same holds for $S_0\exp\big(\rho J_n-\tfrac12\rho^2I_n\big)$.
\end{lemma}

\begin{proof}[Proof of Lemma~\ref{lem:discrete_martingale}] The state $\sigma^N_n$ is a function of $Z^\sigma_0,\dots,Z^\sigma_{n-1}$. Conditional on the normal variables of the first $n$ steps, the ratio $S_{n+1}/S_n$ is therefore log-normal with unit mean, and the conditional expectation of $S_{n+1}$ is $S_n$. The tower property for non-negative variables gives $\E[S_{n+1}]=\E[S_n]=S_0$, so that $S_n$ is integrable. The second statement follows in the same way, with $\rho Z^\sigma_n$ in place of $\rho Z^\sigma_n+\sqrt{1-\rho^2}\,Z^\perp_n$ and $\rho^2(\sigma^N_n)^2h$ in place of $(\sigma^N_n)^2h$. \end{proof}

On every grid with finite adapted volatility the price has expectation $S_0$. For $K^N$ and $\beta>\kappa_2$ its continuous-time counterpart has smaller expectation by Proposition~\ref{prop:finite_converse}. Weak convergence alone would permit this difference because it does not imply uniform integrability.

\subsection{Conditional Monte Carlo under a Gaussian half-tilt}\label{sec:conditional_mc}

Conditional on the volatility normals $Z^\sigma_0,\dots,Z^\sigma_{n-1}$, the log return $X_n-X_0$ of \eqref{eq:price_step} is Gaussian with mean $\rho J_n-\tfrac12I_n$ and variance $(1-\rho^2)I_n$. With $F=S_0$ under our zero-rate convention, a European payoff has the conditional price of Black's formula with forward and total variance
\begin{equation}\label{eq:conditional_forward}
F_n=F\exp\big(\rho J_n-\tfrac12\rho^2I_n\big),\qquad v_n=(1-\rho^2)\,I_n,
\end{equation}
and no draws of $Z^\perp$ are needed \citep{Willard1997,RomanoTouzi1997}. Table~\ref{tab:martingale} reports the sample mean of the raw conditional forward $F_n/F$. We apply neither a multiplicative normalisation by the sample forward nor an additive shift of the prices, so that a deficit of the sample forward stays visible.\footnote{Function \texttt{raw\_forward}.}

For positive correlation the conditional forward has a heavy right tail. In the 15 cells of Table~\ref{tab:martingale} the largest of the 200{,}000 conditional forwards lies between $31$ and $182$ times the forward, and the conditional call price inherits this tail. We therefore price options under a second measure, which we define on the grid by
\begin{equation}\label{eq:half_measure}
\frac{\D\Q^{1/2}}{\D\Q}=\Big(\frac{S_n}{S_0}\Big)^{1/2}e^{I_n/8}=\exp\Big(\tfrac12\sum_{m<n}\sigma^N_m\sqrt h\,\Big(\rho Z^\sigma_m+\sqrt{1-\rho^2}\,Z^\perp_m\Big)-\tfrac18I_n\Big).
\end{equation}

The density has unit expectation by the argument of Lemma~\ref{lem:discrete_martingale}. The superscript $1/2$ is the exponent of $S_n/S_0$ in \eqref{eq:half_measure}, and the share measure of Section~\ref{sec:measure} has the density $S_n/S_0$. \citet{RheinlaenderSchmutz2013} obtain the continuous-time counterpart of the density \eqref{eq:half_measure} as the density of the minimal martingale measure for the log-price. On the grid we call $\Q^{1/2}$ the Gaussian half-tilt. Under $\Q^{1/2}$ the variables $\tilde Z^\sigma_m=Z^\sigma_m-\tfrac12\rho\sigma^N_m\sqrt h$ and $\tilde Z^\perp_m=Z^\perp_m-\tfrac12\sqrt{1-\rho^2}\,\sigma^N_m\sqrt h$ are independent standard normal variables. To simulate under $\Q^{1/2}$ we draw $\tilde Z^\sigma_m$, use $Z^\sigma_m=\tilde Z^\sigma_m+\tfrac12\rho\sigma^N_m\sqrt h$ in the noise map and accumulate $\tilde J_n=\sum_{m<n}\sigma^N_m\sqrt h\,\tilde Z^\sigma_m$. The change of measure is exact on the grid, because $\sigma^N_m$ is known at the beginning of step $m$. In the continuous-time limit the shift adds $\tfrac12\beta\sigma^2$ to the drift of the volatility, so that the quadratic coefficient becomes $\kappa_2-\tfrac12\beta$ (compare Proposition~\ref{prop:closure}). Under $\Q^{1/2}$ and conditional on the volatility normals, the log return $X_n-X_0$ is Gaussian with mean $\rho\tilde J_n$ and variance $v_n$.

For a strike $K$, the identities $(S-K)^+=S-\min(S,K)$ and $(K-S)^+=K-\min(S,K)$ and Lemma~\ref{lem:discrete_martingale} give the call price $C=F-\E[\min(S_n,K)]$ and the put price $P=K-\E[\min(S_n,K)]$. We evaluate the common term under $\Q^{1/2}$,
\begin{equation}\label{eq:half_tilted}
\E\big[\min(S_n,K)\big]=F\,\E^{1/2}\big[\Pi_n\big],\qquad \Pi_n=\sqrt{K/F}\;e^{-I_n/8}\,\Psi\big(\rho\tilde J_n-\ln(K/F),\,v_n\big),
\end{equation}
where $\E^{1/2}$ is the expectation under $\Q^{1/2}$ and
\begin{equation*}
\Psi(z,v)=e^{v/8}\bigg[e^{-z/2}\,\Phi\Big(\frac{z-v/2}{\sqrt v}\Big)+e^{z/2}\,\Phi\Big(-\frac{z+v/2}{\sqrt v}\Big)\bigg],\qquad \Psi(z,0)=e^{-|z|/2},
\end{equation*}
with the standard normal distribution function $\Phi$. Equation \eqref{eq:half_tilted} holds because $\min(S_n,K)$ divided by the density \eqref{eq:half_measure} equals $\sqrt{KF}\,e^{-I_n/8}\exp\big(-\tfrac12|X_n-X_0-\ln(K/F)|\big)$, and $\Psi(z,v)$ is the expectation of $\exp(-\tfrac12|\cdot|)$ under the Gaussian law with mean $z$ and variance $v$. Since $\Psi\le1$, the payoff satisfies $0\le\Pi_n\le\sqrt{K/F}$ for all parameters, and its variance is at most $K/F$. The conditional call price under $\Q$ has no such bound. We use \eqref{eq:half_tilted} under the condition $\kappa_2\ge\beta$ of Theorem~\ref{thm:martingale}, for which the continuous-time price is a true martingale and the relation $C=F-\E[\min(S_T,K)]$ holds in the limit model.\footnote{Function \texttt{conditional\_payoffs}, with the tilt set in \texttt{prepare\_history}. In this subsection $K$ denotes the strike and $K^N$ the kernel.}

\subsection{Kernel and sampling design of the calibration}\label{sec:calibration_design}

The calibration at $H=0.30$ in Section~\ref{sec:illustrations} uses a kernel $K^N$ with $N=97$ positive nodes and weights. The kernel approximates the smoothed kernel
\begin{equation}\label{eq:kernel_smooth}
K_{t_0}(t)=\frac{1}{\Gamma(\alpha)}\Big[(t+t_0)^{\alpha-1}+\frac{b}{\alpha}\,t_0^{\alpha-1}e^{-bt/t_0}\Big],\qquad t_0=\frac{1}{365\cdot512}\ \text{years},
\end{equation}
with $\alpha=H+\tfrac12$. The shift $t_0$ removes the singularity of the fractional kernel at zero. The exponential term restores the mass that the shift removes: its amplitude gives $\int_0^\infty(K_{t_0}-K)\,\D t=0$, and the constant $b=0.0712$ solves $\int_0^\infty(K_{t_0}^2-K^2)\,\D t=0$. We discretise the Laplace representation of the shifted power by Gaussian quadrature with eight nodes on each of twelve intervals, and the exponential term adds one node. At the six maturities of Figure~\ref{fig:motivation} the relative errors of $\int_0^TK^N(t)\,\D t$ and $\int_0^TK^N(t)^2\,\D t$ against the fractional kernel are at most $0.022\%$ and $0.017\%$.\footnote{The script \texttt{rough\_logsv\_table5\_kernel.py} of the replication folder rebuilds the 97 nodes and weights of the file \path{inputs/kernels/H0.3_s512_n8.json}.}

The prices are multilevel estimates \citep{Giles2008} of \eqref{eq:half_tilted} on six grids from 32 to 1,024 steps per day. The coarsest level uses a control variate, the classical model ($H=1/2$) with quadratic drift, simulated by the same scheme at 16 steps per day on the same Brownian increments. A deterministic backward recursion with Fourier inversion gives the expectation of this discrete control. Each level uses 16 independent scrambled Sobol nets with antithetic pairs. The Sobol points drive a Brownian bridge at 16 steps per day, and pseudo-random normal variables, recentred on the bridge increments, fill the finer grids. A net holds from 16{,}384 pairs on the coarsest level to 256 on the finest. The 95\% half-width of a price is $2.131\,\big(\sum_ls_l^2/16\big)^{1/2}+a_{\mathrm c}$. Here $s_l^2$ is the sample variance of the 16 net means of level $l$, and $2.131$ is the $0.975$ quantile of the Student distribution with 15 degrees of freedom. The term $a_{\mathrm c}$ is an allowance for the numerical error of the control mean.\footnote{Class \texttt{RoughLogSVPricer} and function \texttt{sample\_multilevel}.}

\subsection{Moment benchmark}\label{sec:validation}
For linear drift the exact discrete moments obey a linear recursion. The replication code compares these moments with the exact continuous moments for the same kernel and with the sample estimates on each grid.\footnote{The pricing implementation and replication scripts are available in \url{https://github.com/ArturSepp/StochVolModels}.}
\section{Microstructural reading}\label{sec:micro}

Rough Heston and rough Bergomi are scaling limits of order-driven markets. In each case the limit is fixed by two choices, the volatility state that an order moves and the rate at which orders arrive. Table~\ref{tab:kicks} lists the choices. \citet{JaissonRosenbaum2016} and \citet{ElEuchFukasawaRosenbaum2018} start from a Hawkes process in which each trade adds a constant to the trade intensity, which is the variance, and they obtain the square-root noise of rough Heston. \citet{HagerHorstWagenhoferXu2026} start from orders at a constant rate, each of which adds an i.i.d.\ centred impact to the log-volatility, and they obtain the Gaussian log-volatility of rough Bergomi. The candidate construction below uses a constant signed input kick at a rate proportional to the square of its state. The construction identifies the local noise and drift coefficients of our model. Positivity and global convergence require additional arguments. The drift enters through the law of the kicks by construction, and we say so where it matters. The compensated counting process has predictable quadratic-variation rate equal to the order rate. Thus a constant input kick at rate $\sigma^2$ has noise magnitude $\vartheta|\sigma|$, which agrees with the multiplicative coefficient $\vartheta\sigma$ on a positive branch. The kernel subsequently propagates each kick through the state.

\suppressfloats[t]

The sign of the kick decides the drift and the stability condition. A one-signed kick on the variance creates the mean feedback $\varphi*V$ of a Hawkes process with kernel $\varphi$, and stability is the $L^1$ condition $\|\varphi\|_1<1$. The nearly unstable regime $\|\varphi\|_1\to1$ generates the Mittag-Leffler kernel and the mean reversion of rough Heston. A signed kick of mean zero creates no mean feedback, and stability is the $L^2$ condition $\|k\|_2^2<1$ on the kernel $k$ of the quadratic feedback \citep{BlancDonierBouchaud2017}. For linear drift, Proposition~\ref{prop:stationarity} gives the corresponding uniform second-moment condition with $k=\vartheta G$. This condition does not apply to the general quadratic-drift model, for which Propositions~\ref{prop:classical_qd} and~\ref{prop:cesaro} give bounds without a restriction on $\vartheta$. The $L^1$ entry in Table~\ref{tab:kicks} describes the underlying Hawkes feedback kernel. Criticality does not generate the kernel of a signed-kick model. The kernel is an input, and we take it from the empirical decay of the quadratic feedback in \citet{BlancDonierBouchaud2017}, as \citet{GatheralJusselinRosenbaum2020} do.

\subsection{A candidate order-flow model}\label{sec:micro_model}

Let $n$ be the number of trades per unit of annualised variance, so that a trade moves the log-price by one tick of size $n^{-1/2}$. Let $K_n$ be a kernel with $K_n(0)<\infty$ that converges to $K$ in $L^2_{\mathrm{loc}}$, for example the shifted kernel $K_n(t)=(t+1/n)^{H-1/2}/\Gamma(H+1/2)$, which is the kernel of \citet{HagerHorstWagenhoferXu2026} up to normalisation. Trades arrive at the jump times $\tau_k$ of a counting process $N^n$ with intensity $n(\sigma^n_{t-})^2$. Each trade carries a sign $\eta_k\in\{-1,1\}$, which is symmetric, and a volatility mark $\zeta_k\in\{-1,1\}$. The volatility and the martingale part of the log-price are
\begin{equation}\label{eq:micro}
\sigma^n_t=\sigma_0+\frac{\vartheta}{\sqrt n}\sum_{\tau_k\le t}K_n(t-\tau_k)\,\zeta_k,\qquad P^n_t=\frac1{\sqrt n}\sum_{\tau_k\le t}\eta_k .
\end{equation}
The law of the mark depends on the volatility before the trade,
\begin{equation}\label{eq:bias}
\P\big(\zeta_k=\eta_k\mid\F_{\tau_k-},\eta_k\big)=\tfrac12\big(1+\rho(1-|\bar\zeta_n|)+\eta_k\bar\zeta_n\big),\qquad \bar\zeta_n=\mathrm{clip}\bigg(\frac{(\kappa_1+\kappa_2\sigma)(\theta-\sigma)}{\sqrt n\,\vartheta\,\sigma^2}\bigg|_{\sigma=\sigma^n_{\tau_k-}},-1,1\bigg),
\end{equation}
with $\mathrm{clip}(x,-1,1)=\max\{-1,\min\{x,1\}\}$ and with $\bar\zeta_n=0$ defined at $\sigma=0$, where the trade intensity is zero. Then $\E[\zeta_k\mid\F_{\tau_k-}]=\bar\zeta_n$ and $\E[\eta_k\zeta_k\mid\F_{\tau_k-}]=\rho(1-|\bar\zeta_n|)$. The input kick is $\pm\vartheta/\sqrt n$, while the instantaneous state jump is $\pm\vartheta K_n(0)/\sqrt n$. Away from zero the drift enters through a sign bias of order $n^{-1/2}$ per trade.

Let $b(s)=(\kappa_1+\kappa_2s)(\theta-s)$. The compensated input kicks have predictable quadratic-variation rate $(\vartheta^2/n)n\sigma^2=\vartheta^2\sigma^2$, and their mean rate is $\sqrt n\vartheta\sigma^2\bar\zeta_n=b(\sigma)$ wherever clipping is inactive. These two rates are the coefficients entering the kernel convolution in \eqref{eq:model}. In that region the quadratic coefficient contributes the level-independent calming bias $-\kappa_2/(\sqrt n\vartheta)$ per trade, and the other coefficients contribute terms proportional to $1/\sigma$ and $1/\sigma^2$.

For positive $s$, every kick is upward only when $b(s)\ge\sqrt n\vartheta s^2$. If $\kappa_1\theta>0$, this is the interval $0<s\le r_n$, where
\begin{equation*}
r_n=\frac{2\kappa_1\theta}{\kappa_1-\kappa_2\theta+\sqrt{(\kappa_1-\kappa_2\theta)^2+4(\kappa_2+\sqrt n\vartheta)\kappa_1\theta}}.
\end{equation*}
For $K\equiv1$ and $n$ large enough that $r_n>\vartheta/\sqrt n$, the jump chain stays above $\min\{\sigma_0,r_n-\vartheta/\sqrt n\}>0$. This lower bound tends to zero with $n$ and does not prove uniform compact containment away from zero. For the shifted fractional kernel, the state jump is $\vartheta n^{-H}/\Gamma(H+1/2)$, whereas $r_n$ is of order $n^{-1/4}$. When $H<1/4$, a downward jump can exceed the protective interval. Between trades, kernel decay also moves the state. Thus the mark bias does not establish positivity of the fractional construction or prove Conjecture~\ref{conj:wellposed}.

The intensity has the quadratic-feedback form $n(\sigma_0+Z^n)^2$, where $Z^n$ is the kernel-weighted signed input. The state marks and price signs are correlated through $\rho$, and the offset $\sigma_0$ shifts the vertex of the quadratic intensity. These features relate the construction to quadratic Hawkes models \citep{BlancDonierBouchaud2017}, but a fixed positive branch has not been established for singular kernels. With zero initial state and no baseline intensity the process is inactive. A nontrivial comparison with \citet{DandapaniJusselinRosenbaum2021} therefore requires their positive baseline intensity or a different non-zero initial input, as discussed below.

\paragraph{Economic interpretation.}
The construction separates the size of a volatility innovation from the activity that generates it. A trade can increase uncertainty by revealing an imbalance or reduce it by completing an order and absorbing that imbalance. The signed mark represents these two possible effects. We assume that input kicks have the same magnitude at a given scaling level, while trades arrive at a rate proportional to $\sigma^2$. More volatile states then generate more kicks per unit of calendar time, so their aggregate noise amplitude scales with $\sigma$. A larger volatility shock in an active market need not mean a larger input kick from each individual trade. The equal magnitude and the quadratic activity relation are modelling assumptions: the actual state jump also contains the factor $K_n(0)$, and neither the kick size nor the activity relation is estimated in this paper.

The kernel describes how each input is transmitted through time. For $H<1/2$, its power-law weighting can be interpreted as the aggregate effect of participants who revise positions and replenish liquidity over a broad range of horizons. This interpretation motivates the persistence of an innovation without selecting a single reaction time. The power law is imposed on the input kernel. The full response also depends on the drift and the subsequent state-dependent activity. In particular, the linear-drift response is filtered by the resolvent $G$ of Section~\ref{sec:moments}. The analogy with heterogeneous reaction horizons does not derive the fractional exponent from trading behaviour.

Quadratic damping represents a proposed stabilising response that becomes stronger in more volatile states. Above the reference level $\theta$, the drift $(\kappa_1+\kappa_2\sigma)(\theta-\sigma)$ is negative, with leading term $-\kappa_2\sigma^2$ at large $\sigma$. For $\vartheta>0$, the unclipped mean mark in \eqref{eq:bias} approaches $-\kappa_2/(\sqrt n\,\vartheta)$ as $\sigma$ increases: a small, asymptotically level-independent preference for calming kicks combines with the quadratic arrival rate to produce quadratic damping. Inventory adjustment and liquidity replenishment are possible interpretations of this net bias. The mark law is chosen to encode the drift, so this calculation does not independently derive traders' incentives or establish that their response is stabilising in every market.

Positive spot-volatility correlation and a calming average response can coexist. The correlation concerns the joint signs of price and volatility innovations. The mean mark controls the drift. This distinction matters for the pricing model: the stopped share-measure change adds $+\beta\sigma^2$ to the volatility drift, leaving the leading term $-(\kappa_2-\beta)\sigma^2$. Theorem~\ref{thm:martingale} shows that $\kappa_2\ge\beta$ is sufficient for the price to be a true martingale. The inequality is a condition on the model under the specified pricing measure. We do not claim it as an empirical restriction on liquidity providers. The necessity of the condition remains open in the cases stated in Conjecture~\ref{conj:supermartingale}.

\subsection{The Markov case}\label{sec:micro_markov}

\begin{proposition}[Localised Markov approximation]\label{prop:markov}
Let $K_n\equiv K\equiv1$, $\kappa_1>0$, $\theta>0$, $\kappa_2\ge0$ and $\sigma_0>0$. Fix $0<\sigma_-<\sigma_0<\sigma_+$ and $\bar p>0$, and set $D=(\sigma_-,\sigma_+)\times(-\bar p,\bar p)$. The chain $(\sigma^n,P^n)$, started at $(\sigma_0,0)$ and stopped at its first exit from $D$, converges weakly in $D([0,T];\mathbb R^2)$ to the solution of
\begin{equation}\label{eq:markov_limit}
\D\sigma_t=(\kappa_1+\kappa_2\sigma_t)(\theta-\sigma_t)\,\D t+\vartheta\sigma_t\,\D W^\sigma_t,\qquad \D P_t=\sigma_t\,\D W^{(0)}_t,\qquad \D\langle W^{(0)},W^\sigma\rangle_t=\rho\,\D t,
\end{equation}
stopped at its first exit from $D$. The corresponding asset log-price is $X_t=X_0+P_t-I_t/2$.
\end{proposition}

The proof in Appendix~\ref{app:proofs} identifies the generator and proves the stopped diffusion approximation. The proof does not assert global convergence of the chains. Removing the stops requires uniform estimates on their approach to zero and infinity. The candidate limiting diffusion is the classical model of \citet{SeppRakhmonov2023}, and for $\kappa_2>0$ its stationary law and second-moment bounds are given by Proposition~\ref{prop:classical_qd}. This identification explains the role of the calming bias in the candidate limit without supplying a global microscopic limit theorem.

\subsection{Relation to the quadratic Hawkes limit}\label{sec:micro_relation}

Theorem~2 of \citet{DandapaniJusselinRosenbaum2021}, in the numbering of the arXiv version, gives the scaling limit of the purely quadratic Hawkes process with baseline intensity $\mu>0$, Zumbach kernel $k_T=\sqrt{\gamma/T}\,k(\cdot/T)$, $\|k\|_2=1$ and $\gamma\in(0,1)$, in macroscopic time. The limit is $\D P_t=\sqrt{V_t}\,\D B_t$, $V_t=\mu+Z_t^2$ and $Z_t=\sqrt\gamma\int_0^tk(t-s)\sqrt{V_s}\,\D B_s$. Their example takes $k$ proportional to the Mittag-Leffler density $f^{\alpha,\lambda}(t)=\lambda t^{\alpha-1}E_{\alpha,\alpha}(-\lambda t^\alpha)$, which equals the resolvent $R$ of \eqref{eq:resolvent_frac} with $\kappa_1=\lambda$. With $k=G/\|G\|_2$ the limit reads
\begin{equation}\label{eq:djr_limit}
Z_t=\vartheta\int_0^tG(t-s)\sqrt{\mu+Z_s^2}\,\D B_s,\qquad \vartheta=\frac{\sqrt\gamma}{\|G\|_2},
\end{equation}
so that the same parameter $\gamma=\vartheta^2\|G\|_2^2$ controls its stationary second moment. The resolvent identity gives the exact signed-state equation
\begin{equation*}
Z_t=\int_0^tK(t-s)\big[-\kappa_1Z_s\,\D s+\vartheta\sqrt{\mu+Z_s^2}\,\D B_s\big].
\end{equation*}
Formally replacing $\sqrt{\mu+Z^2}$ by $|Z|$ permits the Brownian rotation $\D\widetilde B=\mathrm{sign}_*(Z)\,\D B$, where $\mathrm{sign}_*(0)=1$, and produces a linear multiplicative-noise equation for the signed process $Z$. The rotation does not give that equation for $|Z|$, because taking absolute values does not commute with the Volterra convolution. The rotation also changes correlation with the price Brownian motion according to the sign of $Z$. With $\mu=0$ and the zero initial input of \eqref{eq:djr_limit}, the homogeneous Lipschitz equation has the zero solution. Thus this comparison is a formal analogy for the signed process. The comparison does not identify the limit with our positive-volatility parameter family.

If a stationary second moment exists, It\^o isometry gives $\E[Z^2]=\mu\gamma/(1-\gamma)$. A large ratio as $\gamma\uparrow1$ describes average second-moment amplification, but does not justify omitting the floor pathwise, particularly near zero. Our construction instead inserts the drift intercept $\kappa_1\theta$ through the mark law, and its positivity and convergence questions are separate.

\subsection{The scaling limit for singular kernels}\label{sec:micro_limit}

\begin{remark}[Open problem: a fractional order-flow limit]\label{conj:limit}
For \eqref{eq:micro}--\eqref{eq:bias}, local $L^2$ convergence of $K_n$ alone cannot imply a continuous scaling limit. The state jumps have magnitude $\vartheta K_n(0)/\sqrt n$. For example, the completely monotone kernels
\begin{equation*}
K_n(t)=K(t+1/n)+n e^{-n^3t}
\end{equation*}
converge to $K$ in $L^2_{\mathrm{loc}}$, since the added spike has squared $L^2$ norm $1/(2n)$, but $K_n(0)/\sqrt n\sim\sqrt n$. Before the first trade the state equals $\sigma_0$, so that trade occurs before $T$ with probability $1-e^{-n\sigma_0^2T}\to1$ and has a diverging state-jump magnitude. Such a jump rules out convergence in the Skorohod $J_1$ topology to a continuous process, irrespective of well-posedness of the proposed limit.
\end{remark}

A viable limit theorem must at least impose $K_n(0)/\sqrt n\to0$. Such a theorem also needs uniform kernel regularity sufficient for tightness, moment and exit estimates for the nonlinear marked processes, control of clipping near zero, identification of every limit point and uniqueness of the target equation. One possible kernel bound to investigate is, for some $\varsigma>0$ and each $T$,
\begin{equation*}
\int_0^h K_n(s)^2\,\D s+\int_0^T|K_n(s+h)-K_n(s)|^2\,\D s\le C_T h^{2\varsigma},
\end{equation*}
uniformly in $n$ and $0<h\le1$. This bound is a candidate regularity requirement. We do not prove that it is sufficient for the full nonlinear construction. The shifted fractional example has vanishing jump scale $n^{-H}$, but the sign-preservation difficulty above remains. We therefore leave global convergence to \eqref{eq:model} open. The limits in \citet{DandapaniJusselinRosenbaum2021}, \citet{HorstXuZhang2024} and \citet{HagerHorstWagenhoferXu2026} use their own state spaces, baseline intensities and regularity assumptions, and none is a special case of an established limit theorem for the present construction.

The local drift calculation nevertheless gives a parameter interpretation. The volatility-of-volatility $\vartheta$ scales the signed input kick. The coefficient $\kappa_2$ contributes a level-independent calming bias per trade, and the quadratic arrival rate turns it into quadratic damping. The intercept $\kappa_1\theta$ gives an upward bias at low activity. Under a change of measure, Proposition~\ref{prop:closure} shifts the quadratic coefficient by $\beta\bar\lambda_0+\varepsilon\bar\lambda_1$, subject to the admissibility restrictions stated there.

A bias of order $n^{-1/2}$ does not make the drift unidentifiable from tick data: the relevant limitation is that increasing sampling frequency over a fixed time span does not by itself provide consistent drift estimation, whereas longer observation spans carry drift information. An empirical assessment would require trade-level observations linking arrival rates, signed price moves and volatility innovations, together with an observable proxy for the latent volatility state. The option calibrations in Section~\ref{sec:illustrations} concern risk-neutral coefficients and do not identify this physical order-flow mechanism. We therefore use the construction to interpret the coefficients and establish the localised Markov approximation. A global fractional limit and empirical identification of the trading mechanism remain separate research questions.

\clearpage
\renewcommand{\thetable}{1}
\begin{table}[p]
\tbl{The log-normal Volterra model \eqref{eq:model}, rough Bergomi and quadratic rough Heston with zero variance floor. In quadratic rough Heston, $V_t=a(Z_t-b)^2$ and $Y_t=\sqrt a\,(Z_t-b)$ is the signed state. Its second-moment threshold uses \eqref{eq:gamma} with $\kappa_1=\lambda$ and $\vartheta=\lambda\eta\sqrt a$.}{\begin{tabular}{@{}>{\raggedright\arraybackslash}p{0.15\linewidth}>{\raggedright\arraybackslash}p{0.21\linewidth}>{\raggedright\arraybackslash}p{0.26\linewidth}>{\raggedright\arraybackslash}p{0.28\linewidth}@{}}
\toprule
 & Rough Bergomi & Quadratic rough Heston, $c=0$ & Log-normal Volterra model \\
\midrule
Volatility & Gaussian Volterra process for $\ln\sigma_t$ & $|Y_t|$, with noise $\lambda\eta\sqrt{V_s}\,\D W_s$ in the equation for $Z$ & $\sigma_t$ solves a Volterra equation with noise $\vartheta\sigma_s\,\D W^\sigma_s$ \\
\addlinespace[2pt]
Positivity & Strictly positive for a positive initial variance curve & Non-negative by construction & Non-negative (Theorem~\ref{thm:wellposed}). Strict positivity is open for singular kernels \\
\addlinespace[2pt]
Drift & Deterministic log-variance correction and initial curve & Linear mean reversion of $Y$ to $-\sqrt a\,b$ & Quadratic mean reversion to $\theta>0$ \\
\addlinespace[2pt]
Correlation & Constant $\rho$ & The multiplicative-noise driver of $Y$ has correlation $\mathrm{sign}(Z_t-b)$ with the price driver for $Z_t\ne b$ & Constant $\rho$ \\
\addlinespace[2pt]
Price for $\rho>0$ & Strict local martingale \citep{Gassiat2019} & The correlation depends on the state & True martingale if $\kappa_2\ge\beta$. This condition is necessary for positive finite mixtures (Theorem~\ref{thm:martingale}, Proposition~\ref{prop:finite_converse}) \\
\addlinespace[2pt]
Long-horizon second moment & $\E[\sigma_t^2]$ equals the initial forward variance at $t$ & For $b>0$, $\E[Y_t^2]$ is uniformly bounded if and only if $\gamma<1$ & With linear drift and $\kappa_1>0$, bounded if and only if $\gamma<1$. With $\kappa_2>0$, bounded for every $\vartheta$ and $0<H\le1/2$ (Corollary~\ref{cor:pointwise}). Proposition~\ref{prop:cesaro} also bounds the time average \\
\addlinespace[2pt]
Initial state & Initial forward-variance curve & $Z_0$ through the input curve $\theta_0(t)$ & $\sigma_0$ at $t=0$, with constant input $\Theta_s^0=\sigma_0$ \\
\addlinespace[2pt]
Microstructure & Poisson orders acting on $\ln\sigma$ \citep{HagerHorstWagenhoferXu2026} & Related signed quadratic-Hawkes state \citep{DandapaniJusselinRosenbaum2021} & Marked order flow. The stopped constant-kernel limit is proved. The global rough limit is open \\
\bottomrule
\end{tabular}}
\label{tab:rbergomi}
\end{table}
\clearpage
\renewcommand{\thetable}{2}
\begin{table}[p]
\tbl{Critical volatility-of-volatility $\vartheta_c$ in year$^{-H}$ (left block) and the corresponding one-day innovation proxy $\nu_{1/252,c}$ (right block), with $\kappa_1$ in year$^{-(H+1/2)}$. For linear drift, the second moment is uniformly bounded in time exactly when $\vartheta<\vartheta_c$. The proxy is defined by freezing the volatility coefficient over one day. The closed constant was checked against independent spectral quadrature.}{\begin{tabular}{lrrrrrrrrrrr}
\toprule
 & & \multicolumn{5}{c}{$\vartheta_c$} & \multicolumn{5}{c}{$\nu_{1/252,c}$} \\
\cmidrule(lr){3-7}\cmidrule(lr){8-12}
$H$ & $C(\alpha)$ & $\kappa_1=1$ & $2$ & $3$ & $5$ & $10$ & $\kappa_1=1$ & $2$ & $3$ & $5$ & $10$ \\
\midrule
0.5 & 0.500 & 1.414 & 2.000 & 2.449 & 3.162 & 4.472 & 0.089 & 0.126 & 0.154 & 0.199 & 0.282 \\
0.4 & 0.515 & 1.394 & 1.897 & 2.272 & 2.851 & 3.879 & 0.160 & 0.217 & 0.260 & 0.327 & 0.444 \\
0.3 & 0.574 & 1.319 & 1.711 & 1.992 & 2.413 & 3.129 & 0.279 & 0.361 & 0.421 & 0.509 & 0.660 \\
0.2 & 0.747 & 1.157 & 1.411 & 1.584 & 1.833 & 2.234 & 0.466 & 0.569 & 0.639 & 0.739 & 0.901 \\
0.1 & 1.398 & 0.846 & 0.949 & 1.016 & 1.106 & 1.241 & 0.730 & 0.820 & 0.877 & 0.955 & 1.072 \\
\bottomrule
\end{tabular}}
\label{tab:critical}
\end{table}
\clearpage
\renewcommand{\thetable}{3}
\begin{table}[p]
\tbl{Sample means of the raw conditional forward $F_n/S_0=\exp(\rho J_n-\tfrac12\rho^2I_n)$ of \eqref{eq:conditional_forward} at the last step, $t_n=T=1$, for the scheme in Appendix~\ref{app:simulation} with time step $h$ and the three-factor kernel, with $H=0.1$, $\beta=0.6$, $\varepsilon=0.8$, $\sigma_0=0.5$, $\theta=0.4$ and $\kappa_1=2$. Each cell uses 200,000 paths from seed 3 in 40 batches of 5,000. The coarse grids aggregate the Brownian increments of the 5,760-step grid, and the five columns use the same increments. The expectation of the discrete forward equals one in every cell (Lemma~\ref{lem:discrete_martingale}). The standard errors are the per-path sample value and the standard error of the 40 batch means. For $\kappa_2<\beta$ they describe the sampled paths and give no bound for the distance to the expectation.}{\begin{tabular}{lrrrrr}
\toprule
$\kappa_2$ & 0.0 & 0.3 & 0.6 & 0.9 & 1.2 \\
\midrule
$h=1/1440$ & 0.94109 & 0.98471 & 0.99627 & 0.99858 & 0.99924 \\
per-path standard error & 0.00075 & 0.00155 & 0.00160 & 0.00131 & 0.00112 \\
batch standard error & 0.00073 & 0.00166 & 0.00171 & 0.00134 & 0.00111 \\
$h=1/2880$ & 0.94028 & 0.98516 & 0.99720 & 0.99921 & 0.99967 \\
per-path standard error & 0.00074 & 0.00168 & 0.00183 & 0.00142 & 0.00117 \\
batch standard error & 0.00067 & 0.00180 & 0.00192 & 0.00143 & 0.00115 \\
$h=1/5760$ & 0.93971 & 0.98473 & 0.99711 & 0.99917 & 0.99965 \\
per-path standard error & 0.00072 & 0.00158 & 0.00180 & 0.00143 & 0.00118 \\
batch standard error & 0.00068 & 0.00155 & 0.00177 & 0.00137 & 0.00111 \\
\bottomrule
\end{tabular}}
\label{tab:martingale}
\end{table}
\clearpage
\renewcommand{\thetable}{4}
\begin{table}[p]
\tbl{Fixed-$H$ calibrations to the selected Bitcoin and Ether snapshots, with one six-coefficient vector per column across three maturities. Time is measured in years, and the coefficients use the $H$-dependent units of \eqref{eq:model}. Weighted RMS is the root mean square of the price errors in units of the bid--ask spread, with equal weight for each of the three maturities. The row $\gamma$ gives the stability parameter \eqref{eq:gamma} of the linear-drift model with the fractional kernel at the fitted $H$, $\kappa_1$ and $\vartheta$. Inside-band counts concern premiums. The last two rows report distinct numerical diagnostics in spreads. At $H=0.50$ the error is the largest change of a price across the solver checks. At $H=0.30$ we report the largest pointwise 95\% half-width of the Monte Carlo estimate (Appendix~\ref{sec:calibration_design}). }{\begin{tabular}{lrrrr}\toprule
 & \multicolumn{2}{c}{Bitcoin} & \multicolumn{2}{c}{Ether} \\
 & $H=0.30$ & $H=0.50$ & $H=0.30$ & $H=0.50$ \\
\midrule
$\sigma_0$ & 0.36324 & 0.37746 & 0.77823 & 0.80466 \\
$\theta$ & 0.36317 & 0.36609 & 3.92520 & 2.45560 \\
$\kappa_1$ & 3.80371 & 11.89875 & 0.01003 & 0.57978 \\
$\kappa_2$ & 0.19193 & 0.41383 & 0.57527 & 1.16610 \\
$\beta$ & 0.18193 & 0.40383 & 0.19698 & 0.42839 \\
$\varepsilon$ & 1.87830 & 4.09675 & 0.87174 & 1.87244 \\
\midrule
$\rho$ & 0.09641 & 0.09810 & 0.22041 & 0.22302 \\
$\kappa_2-\beta$ & 0.01000 & 0.01000 & 0.37828 & 0.73771 \\
$\gamma$ at $\kappa_2=0$ & 0.75098 & 0.71211 & 14.47392 & 3.18184 \\
Weighted RMS / spread & 0.71651 & 0.79087 & 0.34923 & 0.30829 \\
Inside bid--ask & 27/45 & 28/45 & 33/39 & 33/39 \\
Max. PDE difference / spread & -- & 0.00116 & -- & 0.00101 \\
Max. price 95\% half-width / spread & 0.04932 & -- & 0.01977 & -- \\
\bottomrule
\end{tabular}}
\label{tab:calibration}
\end{table}
\clearpage
\renewcommand{\thetable}{C1}
\begin{table}[p]
\tbl{The state that an order moves and the order rate fix the noise of the limit model.}{\begin{tabular}{p{2.4cm}p{2.2cm}p{2.6cm}p{1.8cm}p{2.0cm}p{2.95cm}}
\toprule
Limit model & State moved & Kick & Order rate & Noise of the state & Stability norm \\
\midrule
rough Heston & variance $V$ & constant, one-signed & $\propto V$ & $\sqrt V$ & $\|\varphi\|_1<1$ \\
rough Bergomi & $\ln\sigma$ & i.i.d., centred & constant & constant & none \\
model \eqref{eq:model} & volatility $\sigma$ & constant, signed, biased & $\propto\sigma^2$ & $\vartheta\sigma$ & linear drift: $\gamma<1$ \\
\bottomrule
\end{tabular}}
\label{tab:kicks}
\end{table}
\clearpage
\noindent\begin{minipage}{\linewidth}
\renewcommand{\thefigure}{1}\refstepcounter{figure}\label{fig:motivation}
\noindent\textbf{Figure \thefigure}\par\noindent
\includegraphics[width=\textwidth]{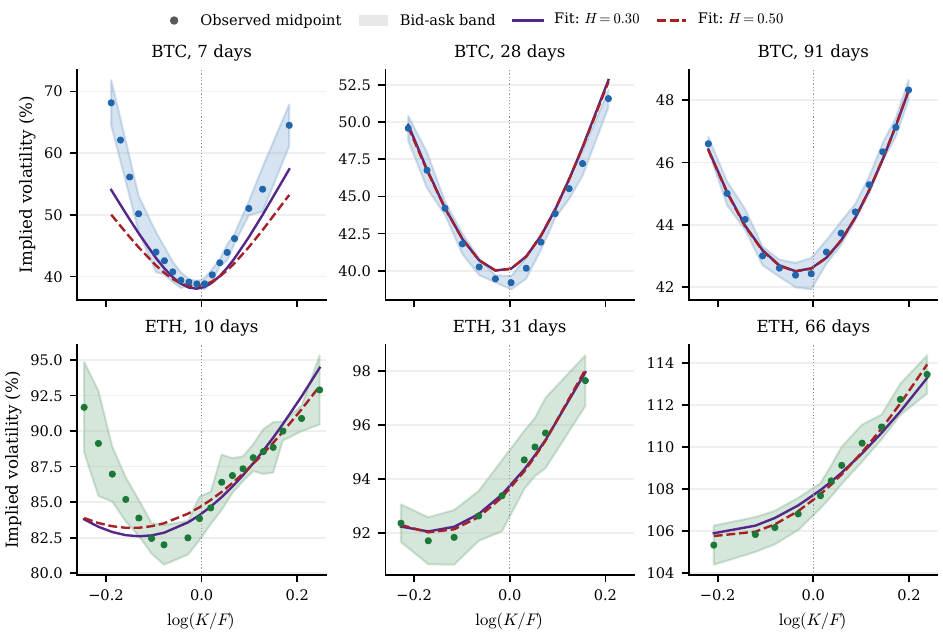}
\end{minipage}
\par\medskip
\noindent\begin{minipage}{\linewidth}
\renewcommand{\thefigure}{2}\refstepcounter{figure}\label{fig:moments}
\noindent\textbf{Figure \thefigure}\par\noindent
\includegraphics[width=\textwidth]{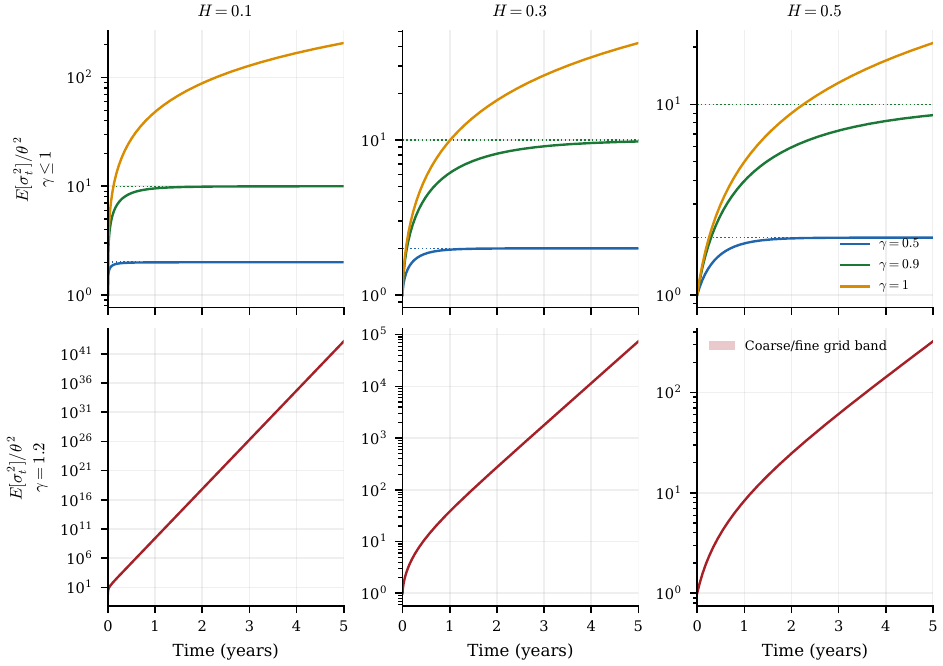}
\end{minipage}
\par\medskip
\noindent\begin{minipage}{\linewidth}
\renewcommand{\thefigure}{3}\refstepcounter{figure}\label{fig:martingale}
\noindent\textbf{Figure \thefigure}\par\noindent
\includegraphics[width=\textwidth]{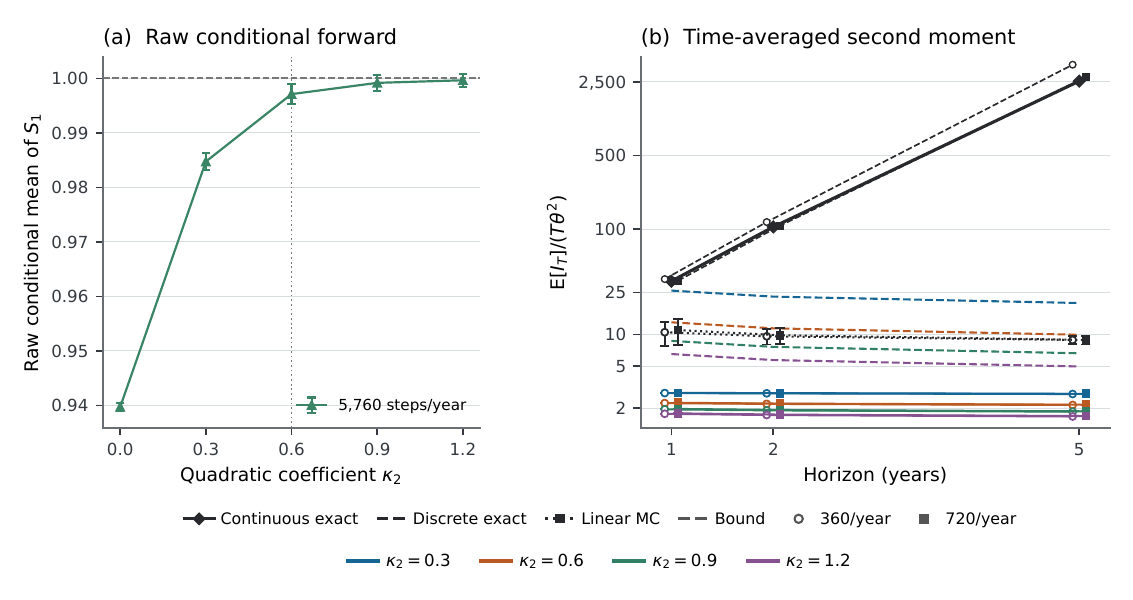}
\end{minipage}
\par
\section*{Figure captions}
\noindent\textbf{Figure 1.} Observed Bitcoin and Ether implied-volatility smiles, with bid--ask bands and model fits for both assets at $H=0.30$ (purple, solid) and $H=0.50$ (red, dashed). Bitcoin is observed on 30 June 2023 at 08:00 UTC with maturities 7, 28 and 91 days, and Ether on 26 October 2021 at 08:00 UTC with maturities 10, 31 and 66 days. The 84 retained out-of-the-money Deribit quotes use one common forward per expiry and unchanged coin premiums inverted through $p=\mathrm{Black}(F,K,T,\mathrm{IV})/F$ for the strike $K$, with unit discounting. Lines connect model values at quoted strikes. Parameters and fit diagnostics are in Table~\ref{tab:calibration}. The cases were selected retrospectively. Data: Deribit quotes archived on 25 October 2026.\par\medskip
\noindent\textbf{Figure 2.} Linear-drift second moments $\E[\sigma_t^2]/\theta^2$ for $\sigma_0=\theta$, $\kappa_1=2$, $\kappa_2=0$ and $H\in\{0.1,0.3,0.5\}$. The top row has $\gamma\in\{0.5,0.9,1\}$, with limits $1/(1-\gamma)$ marked when $\gamma<1$. The bottom row shows $\gamma=1.2$ on separate logarithmic axes. Finite-horizon moments remain finite for every $\gamma$, and the threshold concerns divergence as $t\to\infty$. Shading spans the product-integration solutions on 4,000 and 8,000 steps per year.\par\medskip
\noindent\textbf{Figure 3.} Quadratic-drift diagnostics for the three-factor kernel at $H=0.1$, from the scheme of Appendix~\ref{app:simulation}. Panel (a) shows the sample means of the raw conditional forward of Table~\ref{tab:martingale} at 5,760 steps per year with one per-path standard error. The lines mark a unit forward and $\kappa_2=\beta$. Panel (b) shows the time-averaged second moment $\E[I_T]/(T\theta^2)$ at $T=1,2,5$ years on a logarithmic scale, for $\sigma_0=\theta=0.4$, $\kappa_1=2$, $\rho=0$ and $\vartheta=1.2$ ($\gamma_N=1.0433$, Section~\ref{sec:stability}). Diamonds give the exact linear-drift values in continuous time. The black dashed lines give the exact linear-drift expectations of the scheme at 360 and 720 steps per year. The coloured dashed lines give the bounds \eqref{eq:cesaro} of Proposition~\ref{prop:cesaro}, which hold for every $\vartheta$. Solid lines give the Monte Carlo estimates for $\kappa_2>0$, and the dotted line gives the Monte Carlo estimates for $\kappa_2=0$ as a control. Open circles and filled squares denote 360 and 720 steps per year, horizontally offset, with one standard error from 40 batch means. Each simulated cell uses 200{,}000 paths from seed 3.\par\medskip
\end{document}